REGULAR PAPER

# Checking security compliance between models and code

Katja Tuma[1] · Sven Peldszus[2] · Daniel Strüber[3,4] · Riccardo Scandariato[5] · Jan Jürjens[6,7]

**Abstract**
It is challenging to verify that the planned security mechanisms are actually implemented in the software. In the context of model-based development, the implemented security mechanisms must capture all intended security properties that were considered in the design models. Assuring this compliance manually is labor intensive and can be error-prone. This work introduces the first semi-automatic technique for secure data flow compliance checks between design models and code. We develop heuristic-based *automated mappings* between a design-level model (SecDFD, provided by humans) and a code-level representation (Program Model, automatically extracted from the implementation) in order to guide users in discovering compliance violations, and hence, potential security flaws in the code. These mappings enable an *automated*, and *project-specific* static analysis of the implementation with respect to the desired security properties of the design model. We developed two types of security compliance checks and evaluated the entire approach on open source Java projects.

**Keywords** Security-by-design · Security compliance · Data flow diagram (DFD) · Static program analysis

## 1 Introduction

For decades, organizations have been concerned with the security of their software throughout the entire development process. According to the principle of security by design [18,34,60], the analysis of system assets vis-a-vis security threats needs to be carried out already in the design phase of the development process. In this context, threat analysis techniques (e.g., STRIDE [62], attack trees [59], CORAS [40], and threat patterns [1]) aim to identify security threats to software systems by scrutinizing the architectural design. But, empirical evidence shows that existing threat analysis techniques can be labor intensive [61] and lack automation [66].

Threat analysis is often performed on a graphical representation of the software architecture called *Data Flow Diagram* (DFD) [16,63]. DFD-like models are extensively used in practice, e.g., in the automotive industry [41], at Microsoft [62], and some agile organizations [9]. However, the DFD notation is informal and cannot specify security properties, which are needed to reason about security threats at the design level [23]. To support the detection of problematic information flows at the design level, previous work extends the DFD notation with security-relevant information [8,63] and security semantics [67]. This work uses one such extended notation, namely the Security Data Flow Diagram, in short, *SecDFD* [67] (Sect. 2).

Once a design model has been analyzed and its security flaws fixed, the results are of limited value if the implementation does not comply with the security properties described in the model. Further, design models tend to be useful during the

✉ Katja Tuma
k.tuma@vu.nl

Sven Peldszus
sven.peldszus@rub.de

Daniel Strüber
d.strueber@cs.ru.nl

Riccardo Scandariato
riccardo.scandariato@tuhh.de

Jan Jürjens
juerjens@uni-koblenz.de

1 Vrije Universiteit Amsterdam, Amsterdam, The Netherlands
2 Ruhr University Bochum, Bochum, Germany
3 Radboud University Nijmegen, Nijmegen, The Netherlands
4 Chalmers | Gothenburg University, Göteborg, Sweden
5 Hamburg University of Technology, Hamburg, Germany
6 University of Koblenz-Landau, Mainz, Germany
7 Fraunhofer Institute for Software and Systems Engineering ISST, Dortmund, Germany

design phase but often are ignored after the system is implemented. In particular, empirical evidence shows that only a fraction of open source projects (26% of the investigated projects in [29]) ever update their UML files at least once. Thus, there is a *disconnect* between the architectural design models (containing important security decisions) and the implemented system and its defenses. To be useful for security compliance analyses, an automated connection between the design model and its implementation needs to be established.

Having this connection could also benefit static code analysis. Indeed, existing static analysis tools may report violations that are labeled as *false alarms*, afterward [28]. All reported violations have to be manually sieved through, and, more importantly, the true violations must be distinguished from the false alarms. This is not a trivial task for static program analysis in general, and in particular, it is not trivial for static security analysis (as observed by an industrial experiment in [7]). Making such distinctions can be improved by the contextual information, which can be derived from the (connected) design model.

To address these issues, we propose an approach for security compliance analysis between models and code. This paper is an extended version of our previous conference paper [52]. Specifically, we previously proposed a user-in-the-loop approach (Sect. 3) to support compliance checks between a design-level data flow diagram enriched with security-relevant information (*SecDFD*) and an implementation-level model called *Program Model*, or PM for short (Sect. 2.2). In this work, we contribute with the following extensions:

(i) Two types of security compliance checks using static code analysis (Sect. 4.1).
(ii) An automated extraction of project-specific sources and sinks of confidential information from the design, which we leverage to reduce the number of false alarms raised by an existing data flow analyzer (Sect. 4.2).
(iii) An extended implementation (Sect. 5) including improved user-interface.
(iv) An evaluation of the security compliance checks and data flow analysis with two studies on open source Java projects (Sect. 6).

The two developed types of security compliance checks are relatively precise (average precision is 79.6% and 100%), but may still overlook some implemented information flows (average recall is 65.5% and 94.5%) due to the large gap between the design and implementation. Our approach enables a project-specific analysis with up to 62% fewer false alarms raised by an existing data flow analyzer. These results are certainly valid within the bounds of the threats to validity presented in Sect. 7. We position our contributions in the context of the related work in Sect. 8 and present the concluding remarks in Sect. 9.

## 2 Background

This section describes the background on the design-level model, implementation-level model, architectural compliance, and data flow analysis. We consider the Eclipse Secure Storage [20] to illustrate the models considered in this work. The secure storage allows plugins to store and access secret data. This functionality is used, for example, by the Git extension of Eclipse to store user names and passwords [70].

### 2.1 Design-level model (SecDFD)

At design time, the processing of system data can be specified with a variety of notations. Apart from DFDs, frequently used notations are activity diagrams [14] and business process models (BPMN [6]). Our rationale for focusing on DFDs is twofold: First, they are widely applied in practice, specifically, in the automotive industry [41] and at Microsoft [62] as part of their STRIDE methodology. Second, they represent an essential set of concepts necessary for data flow analysis (processes and data flows between them), which can be mapped exhaustively to activity diagrams and business processes, rendering our mapping generation technique also applicable to these model kinds.

In what follows, we introduce DFDs and the extended security notation which is required for checking the consistency between planned security and implemented security properties.

**Data flow diagrams (DFDs)** A Data Flow Diagram (DFD) is a graphical representation of the software architecture and the information it handles [62]. This type of directed graph represents how the information enters, leaves, and traverses the system. The DFD consists of processes (active entities), external entities (e.g., 3rd parties), data stores (where information rests), data flows (carrying the exchanged information), and trust boundaries (signaling trust levels). Figure 1 depicts a DFD for the Eclipse Secure Storage. The plugin attempts to access a secret by sending a request including path information of where to look for the secret (e.g., a password request for a user name of a Git account). The secure storage queries an internal tree-like data structure to find the corresponding node containing the secret. Next, the cache is queried for the secret value, which can be in clear text (i.e., *secret* on flow 6 in Fig. 1) or encrypted (i.e., *encr. data.* on flow 7). If the value is in clear text, the secret is sent to the plugin. In case of an encrypted value, a decrypt operation either fetches the root password from the operating system or prompts the user to provide it. Upon a successful decryption, the secret is sent to the plugin (flow 10 in Fig. 1). Though useful for performing architectural threat analysis [64], we do not use trust boundaries in our work.

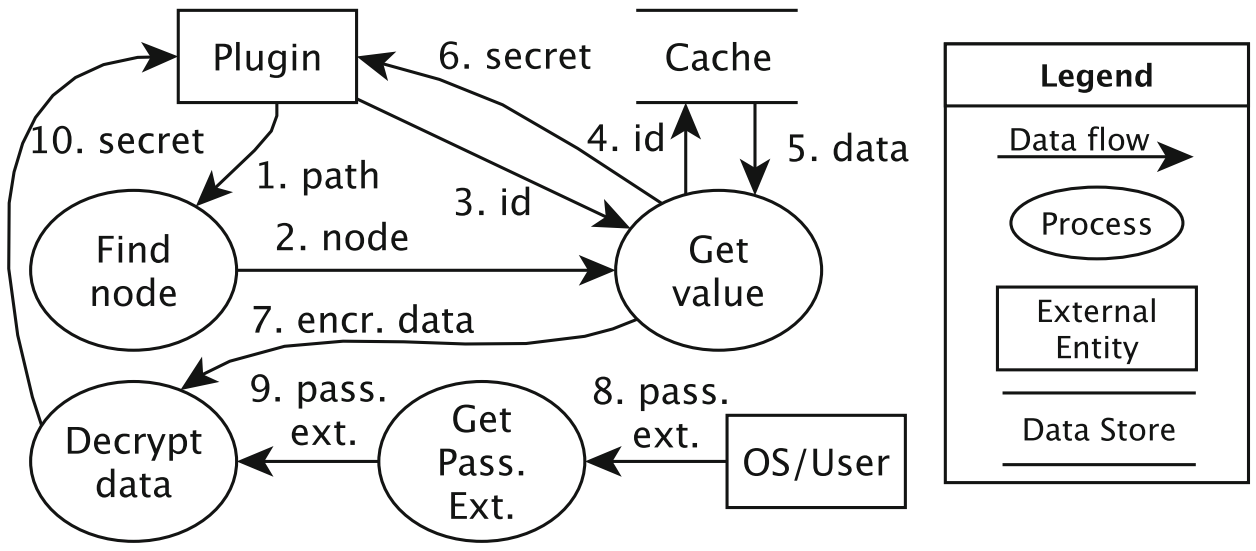

**Fig. 1** A DFD for Eclipse Secure Storage

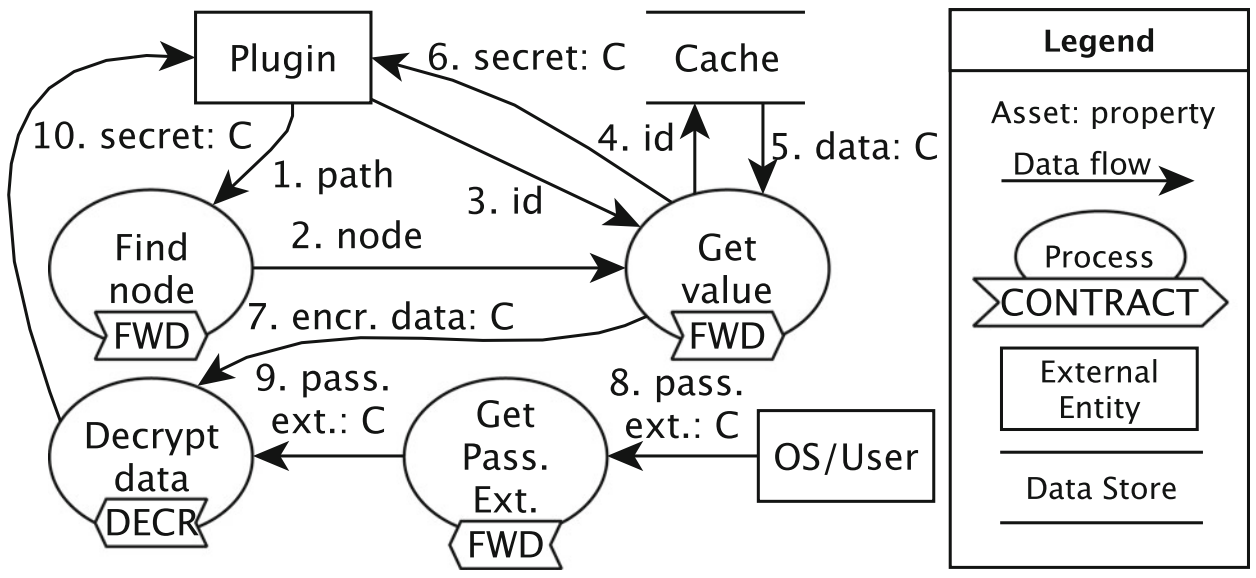

**Fig. 2** An excerpt of the SecDFD for Eclipse Secure Storage

**Security extension** To capture security properties at the architectural level, we use the Security Data Flow Diagram (SecDFD [67]). SecDFD is a notation that enriches DFD with security concepts to enable a formally grounded information flow analysis, focusing on the confidentiality and integrity of information assets. An information asset is specified with a unique name, a source element, (where the asset is first created), target element(s) (where the asset is intended to be consumed or at rest), a type, and a level of confidentiality concern (*high* (annotated with "C" in Fig. 2) or *low*). The data flows that carry assets between other elements (e.g., from an external entity into a process) refer to the specified asset instances. The direction (and order) of the data flows carrying the assets is specified explicitly (as part of the regular DFD notation).

Second, process nodes can be tagged with *security contracts* that define how the security properties of assets change upon exiting the node. A contract of a process node is initialized by specifying the contract type, the incoming and the outgoing assets (which must be transported over some data flows of the process). The contracts are essential for the security analysis with label propagation in the SecDFD. The SecDFD defines four such contracts.

- Encrypt or Hash contract. The contract for encrypting input asset(s) always results in propagating a low (public) label on the output flow(s).
- Decrypt contract. If the input asset is low decrypting, it will result in propagating a low label. However, if the input asset is high decrypting, it will result in propagating a high label on the output flow.
- Join contract. The contract for joining two or more assets propagates the label equivalent to the most restrictive input asset. For example, if a confidential asset is joined with a non-confidential assets, the asset on the output will be confidential.
- Forward or Copy contract. This contract will copy the labels of the input asset(s) to the output flow(s) carrying the corresponding forwarded asset(s).

Finally, the user can specify attacker zones. An attacker zone is defined by a unique name and refers to SecDFD elements (specifically, external entities, processes, data stores, and data flows) which, according to the user, may be part of an attack surface. For the purpose of this study, the attacker zones mark the elements of the model where an attacker is able to *observe* (read) all assets, but other security analyses may require more elaborate attacker models. The user can modify the attacker zones after label propagation and play out several what-if scenarios.

These simple extensions allow us to identify information leaks in the model. For instance, the extended notation [67] is shipped with a simple label propagation (using a dept-first search) according to the specified process contracts. Once the labels have been propagated, a static check is executed to determine if any confidential information flows to an attacker zone.

In comparison with the regular DFD from Figs. 1 and 2 shows an excerpt (for clarity) of the SecDFD for the Eclipse Secure Storage example. For the concrete syntax and semantics of SecDFD, we refer the reader to [67]. If a plugin requires secret data that is cached encrypted, the user must enter a *password* when prompted (c.f. *pass. ext.* in Fig. 2). The externally provided password is then used to decrypt the cached secret data, and if successful the plugin is allowed to read it. First, the designer must specify that the external password is confidential. Second, the designer needs to specify the process contract (e.g., for process *Decrypt_data*). Since the external password is confidential, it should not be leaked to other plugins running in the environment. But, in Fig. 2, the Plugin is not a malicious entity (i.e., it is not part of an attacker zone).

### 2.2 GRaViTY program model (PM)

To create a mapping between SecDFDs and their concrete implementation, we need an easy to analyze representation of the source code. Representations such as abstract syntax trees (AST) contain every detail from the implementation, which makes it hard to analyze for security purposes. Many details about the implementation are not required for our approach. At the same time, important information is not always directly accessible. For example, in the source code files or an AST accesses of fields are not directly visible as access edges between the source and the accessed field, but are access statements within the source to some field with a given name. For our approach, it is only important to know that there is an access to a specific field from some source, but we do not need to know every detail about the circumstances of this access. The Program Model, herein PM (proposed with the *GRaViTY*-framework [48,49,54]), creates a more suitable abstraction for security analysis and allows easy queries, which were very useful for our approach. The *GRaViTY*-framework has been used for the evaluation and execution of refactorings [49], for the detection of anti-patterns [50], as well as for the automated design optimization of Java applications [58].

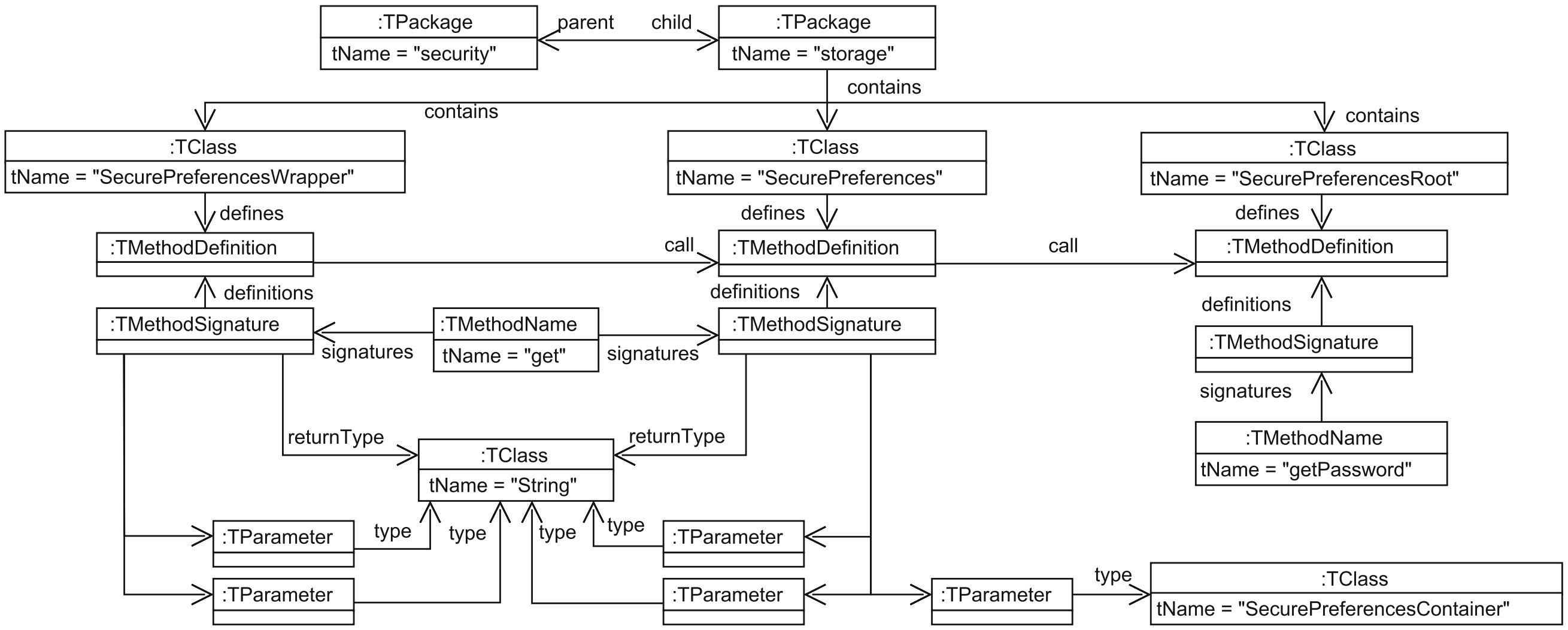

**Fig. 3** Excerpt from the Program Model (PM) of the Eclipse Secure Storage (shown as UML object diagram)

Figure 3 shows an excerpt of the PM created by the *GRaViTY*-framework for the Eclipse Secure Storage example. The figure shows two method calls. The first call is from the method `get(String, String)`, defined in the class `SecurePreferencesWrapper`, to the method `get(String, String, SecurePreferenesContainer)` of the class `SecurePreferences`. The second call is from the called method of the first call to the method `getPassword(String, IPreferencesContainer, boolean)` which is defined in the class `SecurePreferencesRoot`.

On top of the figure, we can see the package structure of the program. All packages without a parent can be taken as entry points for a search. Additionally, it is possible to iterate over all types directly. The types (in this case classes) are shown in the second row, each with a reference to the members defined within the type. For the classes `SecurePreferencesWrapper`, `SecurePreferences`, and `SecurePreferencesRoot`, only a single method is shown. Methods are represented by a triple of method name, method signature and method definition. This allows an efficient search for specific methods, starting with the method name, continuing with signatures, and finding concrete definitions for them. Method signatures have parameters which have a reference to the type representing the parameters type and a reference to the return type. In the PM excerpt, only the parameters of the signature `get(String, String):String` are shown.

A benefit for our mapping from SecDFD to Java implementations is the possibility of an iterative search, starting only with little knowledge about the searched elements—e.g., a method name. The PM allows to start a search with such little information and to find more concrete elements by considering more information like method parameters without iterating over all method definitions defined in the source code.

## 2.3 Compliance

Running compliance checks reveals the relations between a set of components of the design-level model and a set of components of the implementation-level model. As outcome, three different types of relations can be discovered.

**Convergence** In the context of the mappings, convergence means that the user has accepted a suggested mapping or has manually defined a mapping. In the context of security properties, convergence means that a planned security contract is implemented at the correct location and no leaks have been detected by a data flow analyzer.

**Divergence** We refer to divergence when there are flows of assets in the mapped implementation which have not been defined in a DFD. In the context of security properties, we identify divergence when (i) there exists an implemented data flow which does not comply with the specified security contracts of the process node, or (ii) the analysis with a state-of-the-art data flow analyzer reports a leak of potentially confidential information.

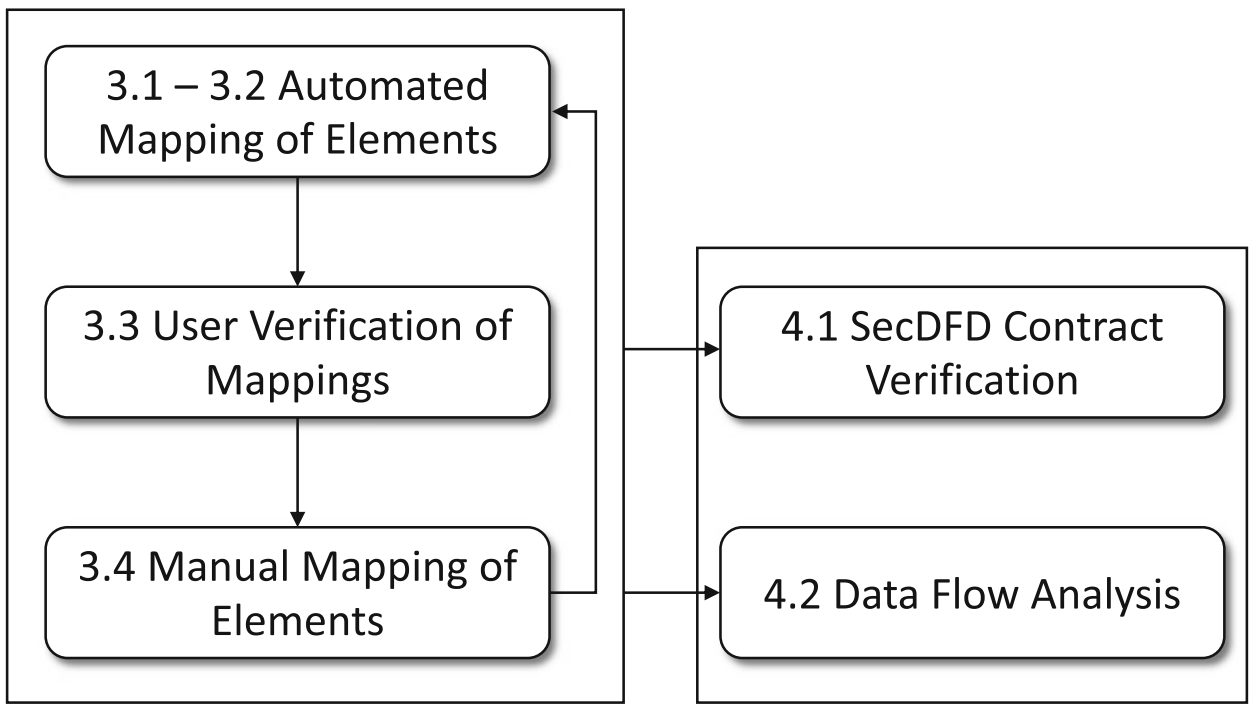

**Fig. 4** Semi-automated mappings approach (Sect. 3) and security compliance checks (Sect. 4)

**Absence** In the context of the mappings, absence means that the user finished using our approach, but there are still design-level elements that have not been mapped. In the context of the security properties, absence means that the SecDFD contracts have not been implemented.

## 3 Enabling compliance checks with automated mapping generation

Assuming a correct DFD, the way it is implemented can vary depending on concrete design (e.g., architectural patterns) and implementation specific decisions (e.g., programming language). Therefore, a fully automatic generation of a correct and complete mapping between DFDs and code is not feasible. Yet, a manual specification of the same mapping is inefficient and error-prone. To this end, we propose an iterative technique for interactively guiding the user in finding an adequate mapping by combining automated mappings with user decisions as shown in Fig. 4. In step 1, mappings between DFD elements and implementation elements are calculated using a heuristic technique. In step 2, these mappings are presented to the user and manually checked by her. In step 3, the user can manually map additional elements. Afterward, the automated mapping is executed again, benefiting from the user input. This process terminates when the user cannot find any additional mapping or finds a violation. Next, the user can perform a security analysis of the SecDFD with respect to the implementation (Sect. 4).

In this section, we describe the steps of our technique in detail, including the automated suggestions. In addition, we explain the use of these mappings for compliance checks. In Sect. 3.1, we define the allowed correspondences between DFD and PM element types. In Sect. 3.2, we show how our automated technique in step 1 establishes concrete mappings between DFDs and their implementations by means of a naming- and structure-based heuristics. In Sects. 3.3 and 3.4, we explain the interactive steps 2 and 3 of our technique. In Sect. 3.5, we argue how the created mappings can be used for checking general compliance.

### 3.1 Corresponding elements

As a prerequisite for mapping DFD elements to code elements, we have to define which DFD element can correspond with which code elements.

**Assets → types** The assets in a DFD are the elements holding critical data. On the level of implementation, data are usually stored in fields, processed using variables and transmitted using parameters and return values. A single asset can be stored in many different locations at the same time which makes it infeasible to map an asset to every single location. The only property of an asset which only changes rarely in programs, written in an object-oriented languages, is the asset type.

**Data stores → types & methods** If we think about data stores like the cache in Figs. 1 and 2, it is quite obvious that this could be a field in some class. But it could also be implemented by an operation which, e.g., requests the cached values from an external server by creating HTTP requests. The common trait between these two variants is the type used to store the data in. The field has a type which provides getters and setters for using the data store, and the method used to get data from a remote server is implemented in a type. Therefore, we map data stores to types as well as to the methods used for accessing the stored data.

**Processes → method(-names)** Processes in DFDs describe functionalities which process data, like methods in implementations do. Obviously, these two elements correspond with each other. While a concrete method definition in an implementation contains all details describing the functionality of this method, the processes only have a name describing the functionality. We assume that a developer implementing a process will choose a similar name for the methods implementing this process. This leads us to a correspondence between the names of processes and the names of methods.

**Processes + Assets → method parameters** Between processes in a DFD, data can be exchanged using flows, where the exchanged data are represented by assets on the flows. In the methods implementing these processes, the same data have to be exchanged. Data between methods in implementations are usually exchanged using parameters and return values. Therefore, we can combine the name mappings between processes and methods with the assets flowing into and out of a process to method parameters giving us the according method signatures.

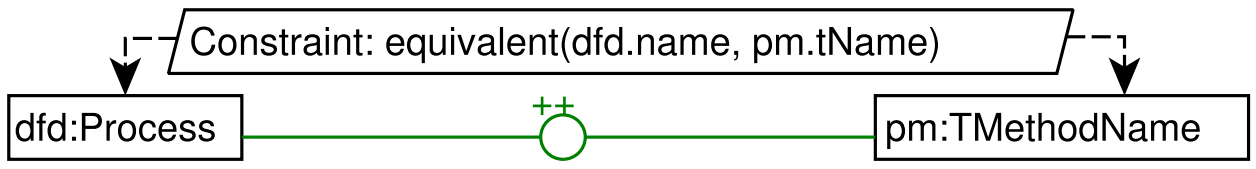

**Fig. 5** Rule describing the name matching for methods

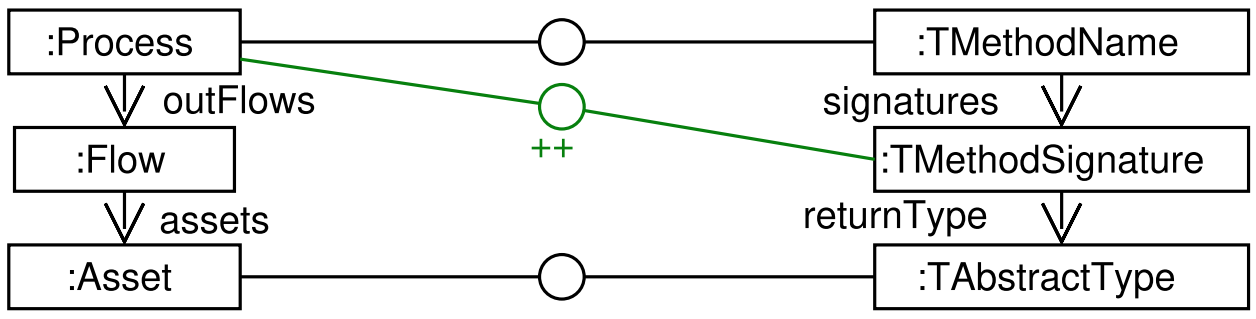

**Fig. 6** Rule for extending name matches based on return types

## 3.2 Semi-automated mapping

In what follows, we discuss the steps of our automatic generation of mappings in detail.

The automated generation of mappings is based on name matchings and structural heuristics, which are sequentially executed and complement each other. For illustration, we formalize two of our mappings using graph rules, using a notation inspired by algebraic graph transformation [21] (explained below). The other mappings can be formalized in a similar way.

**Name matching** First, the names of elements from a DFD are mapped to the corresponding names in the implementation. Asset and data store names are mapped to the names of types, and process names are mapped to the names of methods. Figure 5 shows a rule for mapping processes from a DFD to method names from a PM. A correspondence (visualized as circle connecting the corresponding elements) between a process and a method name is created (denoted by ++) if the constraint at the top of the rule holds. In this case, the names of the two elements on the left and right of the rule have to be equivalent. The precise definition of this equivalence is described in what follows.

Names, both in a DFD and in a Java implementation, are usually built by concatenating multiple words. For example, a Java method name `getPassword` consists of the word *get* and *password*. These words can vary slightly in the names of the corresponding DFD processes (e.g., in plural form, *passwords* instead of *password*). In addition, the style of word concatenation can differ. In Java, usually the camel case (`getPassword`) is used, whereas in DFDs this is not a prescribed style, so underscores may also be used (*Get_Passwords*).

To deal with these issues, we first split the strings at frequently used delimiters and upper-case characters. This gives us for the example the sets of words *[get, Password]* and [Get, Passwords]. Then, we compare the lower-case versions of the words with each other using a fuzzy compare based on the Levenshtein distance [38]. The Levenshtein distance is a measure of the minimal amount of characters which have to be removed, added or flipped to change one word into the other one. For the given example, this distance is zero and one as the first word is already identical and only the character *s* has to be added to change *password* into *passwords*. We accept different distances between words for considering them as identical according to the length of the words to be compared.

Finally, a DFD process is usually implemented in multiple methods, typically having slightly more concrete names. For example besides the method `getPassword`, there might also be an additional method `internalGetPasswod` involved in the implementation of the process *Get_Passwords*. But the DFD process name might also contain additional information—e.g., the process *get_Passwords_External* of the DFD in Fig. 1. To address this challenge, we compare all words from the two names with each other and count the similar words. If this number reaches an threshold of more than half the number of the average words of the compared names, we consider the names sufficiently equal.

For the example DFD in Fig. 1 and the PM excerpt in Fig. 3, we get a name match between the *Get_Value* process and the two method names `get` and `getPassword` as well as a match between the process *Get_Passwords_External* and the method name `getPassword`. While two of this matches are expected, the match between *Get_Value* and `getPassword` is unexpected and should be dropped in the following steps.

**Extending name matches to method signatures** For every method name, multiple signatures may exist. Even if our name matches were always perfectly correct, this would not imply that all signatures with this name are the ones corresponding to a given process. For example, besides the relevant signature `getPassword(String, IPreferencesContainer, boolean):PasswordExt`, there might be a second signature `getPassword():char[]` defined in the Java standard library which is never used in the implementation. To identify the relevant signatures, we use data flow information about assets flowing into and out of a process. Information flowing into a process has to be passed to the implementation of the process, for example, as a parameter value. Likewise, information leaving a process can leave it over return values and parameters. Accordingly, we can use the mapped assets to identify relevant signatures. For every signature, we count how many mapped assets are compatible with the parameters and return types of the existing signatures. If we have at least one match, we consider this signature for further mappings.

A rule for extending a process mapping based on an asset flowing out of a process is shown in Fig. 6. On top of the rule, we can see an existing mapping between a process and a method name, as, e.g., created by the rule shown in Fig. 5. A mapping to one of the signatures having this name is created if there is a mapping between an asset flowing out of the process and a type which is the return type of the signature.

If we look at the return type of the signature `get(String, String):String` and assume that the *secret* asset from Fig. 1 has been mapped to the class `java.lang.String`, we will accept this signature as corresponding with the process *Get_Value*. The other method name corresponding with this process was `getPassword`. The return type of this method signature is `PasswordExt`, and also no parameter type is matching to an asset. Accordingly, we do not create a correspondence.

**Finding implementations of signatures** The last step is to find concrete implementations of a signature corresponding with the process. For every signature, there might be several concrete implementations, all of which do not necessarily correspond to the process. We make use of the flows between different processes to find the concrete definitions.

If there is a flow from one process to another process, this does not only mean that there has to be a signature which has the capability to return or receive the according asset. There also has to be a definition of this signature which is called from a definition in the other process. Therefore, we search for two kinds of data flows between the concrete definitions of the signatures found before.

1. Parameters passed by a call from the source of a flow to to the target of the flow.
2. Return values returned along a call from the target of a flow to the source of the flow.

The flow between two such definitions is not necessarily a single direct call between the two definitions. There can also be multiple definitions in between forwarding data. For example, we can see in Fig. 3 a call between the methods `get(String, String, SecurePreferenesContainer):String` and `getPassword(String, IPreferencesContainer, boolean):PasswordExt` but in the DFD in Fig. 1 there is no flow between the processes *Get_Value* and *Get_Passwords_External*, they have been mapped to. In the implementation, the `get` method forwards the return value of `getPassword` to a call of method `decrypt` which has been mapped to the process *Decrypt_data*. Matching this intermediate to one of the two involved processes is non-trivial. However, if we found such a flow, we can definitely assume that we found two definitions implementing at least parts of the two processes.

The intermediate definitions can be partly mapped to one of the two processes by considering the internal coupling in a process. For every pair of signatures mapped to the same process, we look for pairs of definitions calling each other. For example, this is the case for the definition of the signature `internalGetPassword`, which is called by `getPassword(String, IPreferencesContainer, boolean):PasswordExt`.

**Cleanup** After matching assets and processes, we have to decide which matches are most likely to be correct and, therefore, should be presented to the user. For that reason, we introduce a certainty score for our mappings. This score is calculated with respect to the quality of the underlying name matching as well as the coupling of matched elements with other matched elements. For every DFD element, we only present mappings whose score is higher or equal to the median score of all mappings for this element.

The mappings sorted out in this step are not presented to the user, but may be discovered later again in the interactive process—based on future matches, which might have a coupling to the elements that are now discarded.

### 3.3 User verification of mappings

The mappings created in the previous step are now presented to and verified by the user. For every asset-, data store-type and process-definition mapping, the user can preform three actions.

*Accept:* The user can accept the mapping. From then, the mapping cannot be discarded by the optimization step of the automated mapping approach anymore, and all mappings coupled to this mapping obtain a higher certainty score.

*Reject:* The user can reject the mapping. From then, this mapping is never presented to the user again and it is not considered anymore for extending it to other mappings. All other mappings to which the rejected mapping has been extended will be removed, too, but might be presented to the user again.

*Tolerate:* The user can choose to ignore some suggested mappings. Mappings that are not explicitly accepted or rejected are suggested again and can be re-assessed in future iterations.

Mappings accepted or rejected by the user allow the heuristic to automatically discard related mappings that have only been found by following up the rejected mapping. This is how the search space is reduced in the next automated iteration. Conversely, manually accepting mappings can lead to the score of related mappings being increased and, for this reason, allow to propose new mappings which haven not been considered as correct ones before. Anyhow, a limitation of our heuristic is that they cannot detect mappings which are outside of the search space created by the initial name mappings. We are overcoming this limitation in our approach by including user feedback as described in what follows.

### 3.4 Manual mapping of elements

To increase the search space, an additional user step is conducted after the user manually verified the automatically created mappings (or at least a part of them). In this step, the user has to add at least one new mapping to give additional input to the automated mapping algorithm. The selection of this manually mapped element can have a large impact on the efficiency of the following automated steps.

### 3.5 Compliance of models and code

The mappings can be used to perform compliance checks. In what follows, we describe the check developed to determine if the implementation corresponds with the specification in the DFD.

The correspondence checks take place while the mappings are created. Using the proposed approach, we check for the three kinds of correspondences introduced in Sect. 2.3:
**Convergence** All DFD elements which have been mapped to implementation elements and have not been rejected are allowed to be mapped. Following the definition of convergence, the convergences between the DFDs and the code are described by the set of all allowed mappings.
**Divergence** Elements present in the code, but not specified in the DFD represent a divergence between the DFD and code. To help the user discovering divergences, it is possible to show all flows from members mapped to one process to other members not mapped to this process. If the target of such a flow has not been mapped to any process, there seems to be a divergence. But, a divergence also arises if there is a flow between two processes in the code that has not been specified on the DFD. If an critical asset is communicated along such a flow, this is not only a divergence from the intended design but a security violation.
**Absence** If we are neither able to map a DFD element to the code automatically and the user is not able to map the same element when asked, we discover an absence of specified functionality in the code. Assuming correctness of DFD models, we only have to consider this one direction of the absence (concerning the opposite direction, see *divergence*).

Using these checks, a developer or code reviewer can detect a compliance issue between an DFD and the implementation at hand. However, regarding security, these checks are not precise enough: They might not reveal flows of confidential assets into parts of the program that are not supposed to take place—e.g., if a developer uses a full representation of an object, instead of a stripped one. To this end, we can perform more sophisticated security checks, as described in what follows.

## 4 Security compliance with static program analysis

After the user creates the mappings using our approach (Sect. 3.5), she can use them to verify the security of the implemented systems. First, the developer can automatically verify if the specified SecDFD contracts are implemented (see Sect. 4.1). Second, she can automatically extract project-specific sources and sinks and perform a data flow analysis (see Sect. 4.2). The provided feedback of compliance violations and potential leaks may cause her to revisit the implementation, and reflect the changes in the SecDFD.

### 4.1 Verification of specified SecDFD contracts

This work used *taint analysis* techniques, which are a kind of information flow analysis techniques where data objects are tainted at the source and tracked to the sink using data flow analysis [39]. Concretely, we developed static checks to verify the compliance of the implementation to the SecDFD encrypt, decrypt, forward, and join contracts. We assume an existing mapping between the SecDFD and the implementation before executing the checks.
**Encrypt and decrypt contracts** When executed, all encrypt and decrypt process contracts will be checked against the implementation. For each process with such a contract, we collect all the mapped method implementations that call at least one method signature performing an encrypt or decrypt operation. If at least one such method implementation exists, we consider that the process contract has been implemented, and mark it as *convergence*. If no such method implementation has been mapped to this process, we consider that the process contract has not been implemented, and mark this occurrence as *absence*.

We provide a list of well-known methods that are called during cryptographic operations. We compiled this list by inspecting the Java standard security library, and packaged it together with the plugin. In addition, the user is able to add project-specific methods to this list (at runtime) via the user interface. We remark that state-of-the-art static analysis tools (e.g., SonarCube[1]) maintain similar rules for checking implemented encryption logic, but with our approach users can verify their expectation regarding the planned security.
**Forward and join contracts** The forward and join contracts at the SecDFD level describe local data flows within a process that have to be present in the implementation. To check if the specified contracts have been implemented, we propose a two-step procedure introduced in what follows. First, we extract the relevant asset-communicating flows from the process's implementation (I-Flows). Second, we compare the implemented flows with the expected flows specified in the SecDFD (D-Flows).

The main challenges in checking forward and join contracts are that one process can be realized by multiple methods, but there are also many methods that do not belong to any process but interact with multiple processes. Further-

1 https://www.sonarqube.org.

more, an asset in the SecDFD can be realized by different types in the implementation. For example, the encrypted data (encr. data) in Fig. 2 are realized by instances of the Java classes `String` and `CryptoData`. In addition, a single type in the implementation can be used to create instances of different assets. This is especially a problem for frequently used types like strings that can be used to represent nearly every asset as shown before.

```
Input  : Process p, Mapping m
Output: I-Flows i
1  methods ← m.methods(p)
2  in ← inFlows(methods)
3  foreach flow ∈ in do
4      type ← communicatedType(flow)
5      if m.mapping(type) = ∅ then
6          remove flow from in
7      end
8  end
9  out ← outFlows(methods)
10 foreach flow ∈ out do
11     type ← communicatedType(flow)
12     if m.mapping(type) = ∅ then
13         remove flow from out
14     end
15 end
16 i ← {}
17 foreach target ∈ out do
18     sources ← reachableBwd(target, in)
19     if sources ≠ ∅ then
20         add (sources,target) to i
21     end
22 end
23 return i
```

**Algorithm 1:** Algorithm for the Extraction of the I-Flows *i* for a given Process *p*

In Algorithm 1, we show the pseudocode for the extraction of the implemented flows (I-Flows) for a given process. For this purpose, we leverage a data flow graph as it is constructed by state-of-the-art tools such as Soot [12]. To simplify the implementation, we used a preliminary version of an extracted data flow graph at the abstraction level of the PM [42] that provides an immediate coupling among PM elements, such as parameters, and the data flow graph. Following the definition of the forward and join contracts, we search for implemented data flows from one or more source methods that flow through the methods implementing the process into a single target method. Accordingly, an I-Flow object consists of the flow's target method and a set of the source methods from which the data flow originates. The inputs to this algorithm are the process for which we want to extract the implemented flows and the mapping described in Sect. 3.1.

First, in line 1, we retrieve the methods implementing the process *p* from the mapping *m*. For these methods, we search in line 2 for the relevant incoming data flows in the implementation. To this aim, we implement the operation *inFlows* that returns all data flows entering a set of methods. This operation aggregates all incoming data flows into the methods' parameters and all return flows of called methods.

Next, we filter the collected data flows in lines 3–8. For the forward and join check, only the flows that can be used to communicate assets from the SecDFD are relevant. This means that the type communicated along a data flow has to be mapped to an asset. Accordingly, we filter out the flows which communicate unmapped types. At this point, it is not important which assets can be communicated along the single data flow.

In lines 9–15, we proceed similarly for the data flows leaving the methods implementing the process. Comparable to *inFlows*, *outFlows* collects all data flows leaving the set of methods. These are the return flows of the methods themselves and all data flows from the methods into parameters. Again, we filter the flows according to the communicated types.

After filtering, in line 18, we determine for every outgoing flow (*target*) which of the flows entering the process's implementation (*in*) can flow into this outgoing flow using a backward search on the data flow graph. In line 19, we check if we found reachable incoming flows (*sources*). The pair of the found sources and the target represent one I-Flow, that is added to the result set *i*. If exactly one incoming data flow is propagated to the outgoing data flow, we found an *implemented forward*, and if multiple incoming data flows are propagated to an outgoing data flow, we found an *implemented join*. Note that we only consider patterns with one outgoing flow. If there are contracts in the DFD with multiple outgoing flows, they have to be split into multiple contracts. Finally, we return all found I-Flows.

After we extracted the I-Flows, we compare them to the expectations from the SecDFD using Algorithm 2. The input to this algorithm is the process, the mapping, and the extracted I-Flows. The output is a set of identified violations (absence and divergence).

The algorithm is again based on two steps. First, we collect all possible matches between the I-Flows and the expected flows from the SecDFD contracts (D-Flows). We consider the implementation of a contract to be *convergent* with the SecDFD if and only if there exists a bidirectional one-to-one mapping between every D-Flow of all contracts and all I-Flows extracted using Algorithm 2. We call this property a biunique mapping. But, the matches are usually not biunique because of the overlapping asset type mappings; therefore, we have to reduce the initial set of matches to a set of biunique mappings in the second step.

To collect the matches, we iterate over every contract and every outgoing asset of the contract in lines 2 and 5. For each of these pairs, we select I-Flows if their possible outgoing

```
Input  : I-Flows i, Process p, Mapping m
Output: Violations v
1 v ← {}
2 matches ← {}
3 foreach contract ∈ fwdJoinContracts(p) do
4     inAssets ← contract.inAssets()
5     foreach outAsset ∈ contract.outAssets() do
6         flows ← {}
7         foreach iflow ∈ i do
8             type ← communicatedType(iflow.trg())
9             if outAsset ∈ m.mapping(type) and ∀ s ∈ iflow.src() :
              (m.mapping( communicatedType(s)) ∩ inAssets) ≠ ∅
              then
10                add iflow to flows
11            end
12        end
13        if flows = ∅ then
14            add "Absence: Not implemented" to v
15        end
16        add (contract, outAsset)→flows to matches
17    end
18 end
19 solution ← findSolution(matches)
20 if solution = ∅ then
21     add "Divergence: No biuniqe assignment" to v
22 else
23     foreach flow ∈ (matches \ solution.flows()) do
24         add "Divergence: Not in DFD" to v
25     end
26 end
27 return v
```

**Algorithm 2:** Algorithm Checking the Implemented Flows *i* for a given Process *p* against the Specified Contracts

assets contain the expected asset and if for every incoming flow at least one possible asset is contained in the set of expected incoming assets (see line 9 in Algorithm 2). If no such I-Flow exists, the contract is not implemented (for this outgoing asset) and we detect an *absence* (lines 13 and 14).

After collecting all possible matches, we have to find a biunique solution within the created mappings between the D-Flows and the I-Flows. This is implemented in the function *findSolution*. The easiest implementation is to iteratively assign I-Flows to D-Flows and to check if a solution is still possible, meaning that there is no unassigned I-Flow that cannot be assigned to an unassigned D-Flow anymore. If so, we can assign the next I-Flow to a D-Flow, else, we have to backtrack. If we cannot find such a solution, we report a violation as there is at least one not implemented contract and we detected an *absence* (lines 20 and 21). If we found a solution, all specified contracts have been implemented and we found a *convergence*. However, all I-Flows that are not part of the solution are still reported as violation as they are unspecified forwards or joins of assets and represent a *divergence*.

## 4.2 Optimized data flow analysis

To perform a data flow analysis, the developer needs to identify the sources and sinks of secret data in the implementation. More importantly, to perform a meaningful and precise data flow analysis, the sources and sinks must be *identified correctly*. For instance, we have found the standard substring method in Java (`java.lang.String.substring (int, int):String`) as one of the sink method signatures in an existing list of identified sinks.[2] This will result in many false alarms raised by the analyzer, since it seems unlikely that data can leave the system through this method and it is a very common operation over strings in Java. Dually, overlooking an important source may result in overlooking true leaks. Though some sources and sinks can be extracted from library APIs [4,56], finding project-specific sources still remains a challenge. In addition, many data flow analyzers work with a flat security policy. Specifically, they raise an alarm if there is an access path between *any* of the source methods and *any* of the sink methods. But, certain tainted data might be expected to flow to some sinks (e.g., writing an encrypted password to local storage) but not others. If all the tainted objects are treated equally, the analyzer raises false alarms. In response to this challenge, we aim to automatically extract project-specific sources and sinks *for each SecDFD asset*.

**Project-specific sources** The SecDFD requires the user to specify confidential assets, thus their source element (in the model) can easily be determined. There are three possible types of source elements: an external entity, a data base, or a process. If the asset source is an *external entity* and it is mapped to method definitions, their signatures are collected as sources. But, if a mapping of the external entity does not exist (e.g., for the entity Plugin from Figs. 1 and 2), the signatures of the mapped method definitions of the processes reading from that entity are collected instead. If the asset source is a *data store*, it can be mapped to methods or types. First, the signatures of method definitions mapped to the data store (if any) are collected. Second, if the data store is mapped to a type (e.g., a Class), the signatures of method definitions defined by this class are also collected, but only if the return type matches the asset type. Finally, an asset source can be a *process* element (e.g., a random number generator). If there is no process contract with this particular asset on the output, then the signatures of the method definitions mapped to the process are collected. But, the asset may originate in the process as a result of a transformation (e.g., a join of two assets). In this case, the assets on the contract inputs are *traced backwards* reaching either an external entity, a data store, or a process with no contracts impacting the traced asset. The signatures of the method definitions mapped to the traced element are collected as sources.

[2] https://github.com/secure-software-engineering/SuSi.

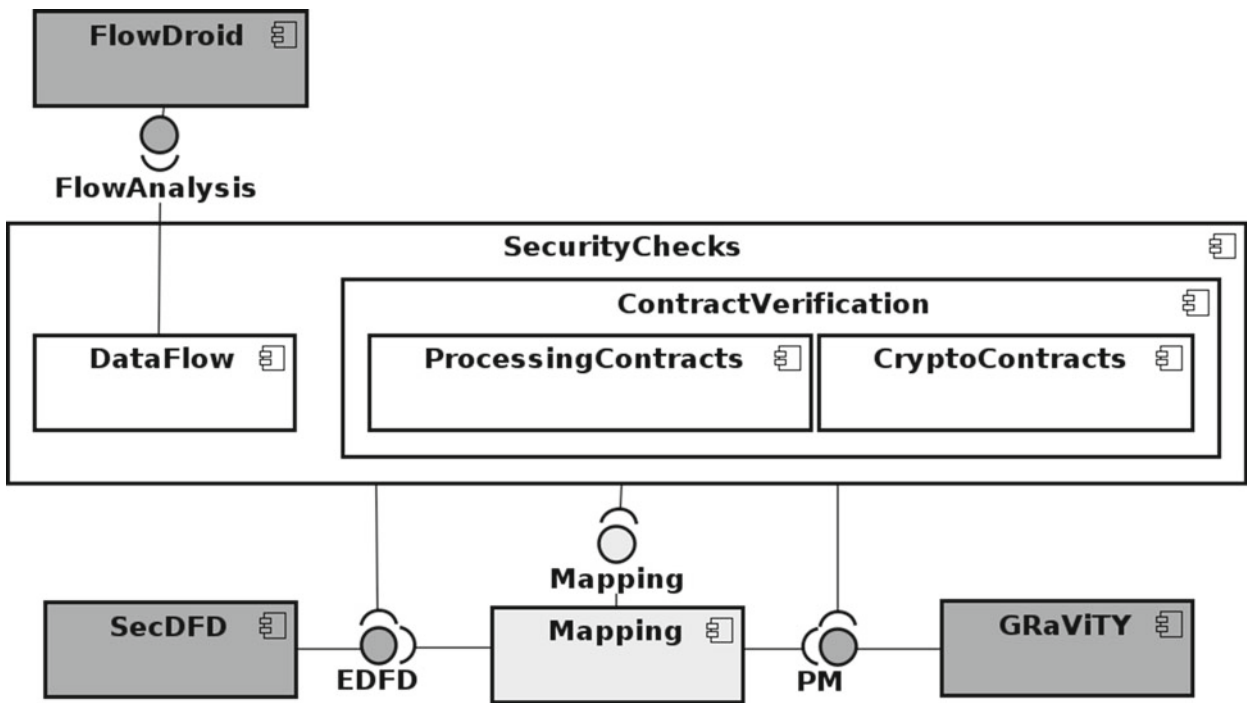

**Fig. 7** Architecture of the implementation

**Allowed sinks** We collect the sink method signatures from [4,56] (excluding methods of Android specific packages) and exclude the allowed sinks. The allowed sinks are maintained *for each* confidential asset. These are method implementations mapped to SecDFD elements where the confidential asset exits the system (i.e., external entities and data stores). For example, the secret flowing into the Plugin (data flow 10 in Fig. 2) is expected to flow there. Therefore, we consider the Plugin as an allowed sink. However, since the Plugin cannot be mapped to the implementation, we instead consider the method implementations mapped to the Decrypt data process as allowed sinks.

**Attacker zones** The SecDFD allows the user to specify attacker zones, which denote what elements are observable by the attacker. For each asset, we collect signatures of all the method definitions mapped to elements of attacker zones and add them to the list of sinks and (if needed) remove them from the allowed sinks. In this way, the user is able to influence the security policy of the SecDFD and perform an analysis assuming over-exposed components or APIs. This kind of what-if analysis can be useful to identify the impact of a security mitigation on the design level.

## 5 Implementation

In this section, we give a quick overview of the implemented tool. We refer the interested reader on how to use the tool to "Appendix". The approach is implemented and packaged as a publicly available Eclipse plugin [53]. The architecture of our implementation is shown in Fig. 7. Our implementation is structured according to the two main contributions of this work. First, we have the semi-automated creation of mappings realized in the component `Mapping`, and second, the security compliance checks realized in the `SecurityChecks` component.

**Semi-automated mappings** For the creation of mapping suggestions, we implemented the name matches and the patterns shown in Sect. 3 in hand written-java code. The implementation leverages an existing implementation for modeling SecDFDs using an Xtext DSL with editor support [67]. Also, we use an existing plugin for generating the Program Model from Java source code [54]. The SecDFD and the PM are accessed through the Java APIs provided by the components realizing the models.

Using the Eclipse API, an integration into the Java source code editor is provided. For working with the SecDFD, the textual editor is provided by the `SecDFD` component. In addition, we provide a graphical Editor based on the Sirius framework.[3] For showing the proposed mappings to the user, we registered a view in the Eclipse IDE. As a single system is usually described in multiple SecDFDs, we extended the implementation of this view to support multiple SecDFDs at the time. Details on how the user interacts with our implementation are presented in "Appendix". Created mappings can be accessed through a wizard that shows all SecDFDs within a project as well as all existing mappings.

Access to the mappings is provided to other components though a `Mapping` interface, e.g., for the verification of SecDFD contracts. This interface allows to query the mappings in both directions, for mappings to a given SecDFD element and mappings to PM elements for a single or multiple SecDFDs.

**Security checks** The implementation of the security compliance checks is following the structure of Sect. 4 and is separated into two components. One component for performing optimized data flow analyses (`DataFlow`) and one for the verification of the contracts specified in a SecDFD (`ContractVerification`).

In this work, we perform the data flow analysis using FlowDroid [5], a state-of-the-art taint analyzer for Android applications, but also applicable to Java programs. The 2.7.1 release of FlowDroid was obtained from its release website[4] and is imported as a library in our plugin.

FlowDroid raises an alarm if and only if an object flows from a predefined list of *source* methods (i.e., these objects are tainted) into *sink* methods (i.e., they violate the security policy). The sources and sinks must be identified and are passed as parameters to the analyzer. To simply the analysis, FlowDroid relies on capabilities of the Soot compiler framework [36] which converts Java bytecode into the Jimple [68] intermediate code representation. This makes the analysis in FlowDroid precise as it is flow-sensitive (the call graph is aware of the order of statements) and context-sensitive (the call graph is enriched with the context of the callees).

[3] https://www.eclipse.org/sirius/.

[4] FlowDroid Release Site: https://github.com/secure-software-engineering/FlowDroid/releases.

In addition, the Jimple representation is able to handle Java reflection, but only for reflective calls where the types of all referenced classes are known. The analysis in FlowDroid is also object-sensitive (i.e., the call graph distinguishes method invocations on different object instances) since it uses access paths as taint abstractions. In general, taint analyzers consider only explicit flows for performance reasons [26], but FlowDroid also supports tracking implicit flows and shows high performance results on benchmarks (86% precision and 93% recall on DroidBench [5]). We refer the interested reader to [3] for more details. The `DataFlow` component of our implementation executes FlowDroid over its Java API. Following Sect. 4.2, we execute FlowDroid for every asset in the SecDFD taking its set of allowed sinks and possible sources into account.

The contract verification is again split into two subcomponents: one for the verification of the forward and join contracts (`ProcessingContracts`) and one for the verification of the encrypt and decrypt contracts (`CryptoContracts`). In both sub-components, we implemented the checks as introduced in Sect. 4.1 using handwritten Java code.

## 6 Evaluation

The evaluation of our approach has three parts. First, we conducted an experiment to evaluate the automated mapping creation (Sect. 6.1). Next, we conducted experiments to evaluate the verification of SecDFD contracts in implementation (Sect. 6.2), and the optimization of data flow analysis by extracting project-specific sources and sinks from SecDFDs (Sect. 6.3).

Table 1 depicts the characteristics of five open source Java projects used in our studies.

*Jpetstore* [45]. This is a web application built on top of MyBatis 3, Spring and the Stripes Framework. This is an example with very few classes, implementing the basic functionalities of a web store. In principle, the users are able to create their accounts, browse, and order goods online. Jpetstore has been designed as minimal demonstration application for MyBatis, which should have a good design and documentation. The developers tried to strictly follow the MVC pattern.

*ATM simulation* [11]. This is a simulation for an ATM machine developed for academic purposes. The ATM simulation implements the main procedure of a control system. Upon start-up, a new session is initiated, and the users are able to insert their card and PIN number. The session continues upon a correct PIN entry and provides the users with the option of a withdrawal, deposit, balance inquiry, and money transfer. After a completion of desired transactions, the ATM returns the card and optionally prints the receipt.

*Eclipse Secure Storage* [20]. As described in Sect. 2, Eclipse Secure Storage is used for ensuring secure storage and management of sensitive data within the developer's Eclipse workspace. The secure storage allows for plugins to authenticate and have controlled access to workspace resources.

*CoCoME* [33]. CoCoMe is a platform for collaborative empirical research on information system evolution [30]. This platform helps engineers manage different aspect of software evolution, such as the system life-cycle, versioning artifacts, and comprehensive evolution scenarios. The implemented system is a cash register.

*iTrust* [43]. This example is a web application for hospitals which allows the hospital's staff to manage medical records of patients, based on 55 use cases. The example originally stems from a course project has been maintained by the Realsearch research group at North Carolina State University and was used as an evaluation example in research papers before [13]. Detailed requirements describing different activities are available [43]. However, the available requirements and use cases mostly describe very simple tasks and only a few of them are realized in the implementation.

### 6.1 Evaluation of mappings

The purpose of this study was to evaluate the correctness of the suggested mappings. In what follows, we briefly describe the design of the experiment, the projects, and the results.

**Design of study** We conduct this experiment with all five open source projects from Table 1. To evaluate the correctness of the suggested mappings, we set up an experiment to compare a ground truth of manually created mappings with the generated mappings for each of the five considered projects. The iterative approach involves the user to guide the generation of mappings in the desired direction. As per this design choice, we intentionally investigate the correctness of the automated mappings and the impact of the user separately. Consequently, the evaluation aims to answer the following research questions.

*RQ1. What is the correctness of the automated mappings generated by the plugin?* We measured correctness in terms of *precision* and *recall* (dependent variables). Conventionally, precision $(TP/(TP + FP))$ is measured as a ratio between the true positives (i.e., correct mappings) and all generated mappings (including the false mappings). A true positive $TP$ is a correct mapping between the source code and the DFD element which is listed in the ground truth. A false positive $FP$ is a mapping between the source code and DFD element that is not listed in the ground truth. Recall $(TP/(TP + FN))$ is measured as a ratio between the true positives and all correct mappings (including the overlooked mappings). A false negative $FN$ is a mapping between the source code and the SecDFD element which is present in the ground truth, but has not been identified.

**Table 1** Projects considered in the evaluation

| Project | Source code | | | DFD |
|---|---|---|---|---|
| | Lloc | Classes | Methods | Elements |
| jpetstore | 1221 | 17 | 277 | 47 |
| ATM simulation | 2290 | 57 | 225 | 85 |
| Eclipse Secure Storage | 2900 | 39 | 330 | 41 |
| CoCoME | 4786 | 120 | 512 | 44 |
| iTrust | 28, 133 | 423 | 3691 | 31 |

*RQ2. What is the impact of the user on the correctness of mappings?* The implementation automatically derives trivial mappings from the user defined mappings, raising the recall before a new iteration starts. Therefore, the impact of the user defined mappings is measured as the difference in recall before, and after the added mappings.

**Execution** The experiment was executed by the first and second author. The authors worked on the projects individually and compared their results at each step. First, the authors created the SecDFDs for all five projects models manually. To this aim, the authors inspected all available documentation (including the source code) and reverse engineered a high-level architecture. Second, a ground truth was created for each SecDFD by following the execution of the modeled scenarios and manually mapping the executed methods and transferred data to the processes and assets of the according step. The ground truth is a JSON file with a list of correspondence mappings between the elements of the SecDFD and a uniquely identifiable location of the source code element. Third, the implemented plugin was used to find the automated mappings in several iterations. Each iteration included accepting, rejecting the automated mappings, and defining mappings manually by highlighting elements in the source code and specifying the corresponding SecDFD elements. After each iteration, the precision and recall of the automated mappings were logged.

**Results** This study shows promising results for guiding the user in the discovery of compliance violations.

In particular, Table 2 shows measurements of high precision and recall only after a few iterations for realistic Java projects. Each iteration consists of an automated, and a manual (user input) phase. We present the precision and recall for the automatically suggested mappings in each iteration. We also depict the amount of manually accepted, user defined, the sum of all accepted and user defined, rejected mappings, and the impact of the user defined mappings on recall (in that order). Notice that the later iterations make use of the manually defined mappings.

*RQ1.* We start by reporting the correctness of the automated mappings in the first iteration. The average precision of the first iteration is 50.5%. On average, the recall of the first iteration is 69.8%. Yet, both the precision and the recall increase after the first iteration. On average, the final precision and recall of the automated phase are very good (87.2% and 92%, respectively).

The average difference between the recall of the second iteration and the the user-impacted recall of the first iteration (last column in Table 2) is 4.5%. This means that on average, the automated search was able to increase the recall between the first and second iteration by 4.5%.

On the other hand, the average difference between the user-impacted recall of the second iteration and the recall of the third iteration is minimal. This means that, the automated search was not able to increase the recall significantly between the second and third iteration.

*RQ2.* On average, the user accepted less (7) mappings then they rejected (9.6) and defined only 2.6 mappings manually. However, in three cases (jpetstore, ATM simulation, Eclipse Secure Storage) the user accepted more mappings then rejected. This means that the user could quickly scan the suggested mappings and eliminate the ones that are obviously wrong. Overall, adding a few mappings manually resulted in a more fruitful next iteration. For instance, adding three mappings manually in the first iteration of evaluating the ATM simulation resulted in two new correct mappings (see accepted mappings of the second iteration).

On average, the user impact on the recall was an increase of 7.9%. This means that the users were indeed able to guide the discovery of compliance violations. Further, the users had a larger impact on increasing the recall in later iterations compared to the automated search (7.9% vs 4.5%). Notice, that on average 75% of all correct mappings ($TP$) are suggested to the user and do not have to be manually defined.

## 6.2 Evaluation of the SecDFD contract verification

In this section, we evaluate if the proposed contract checks (Sect. 4.1) can effectively detect convergence, absence and divergence between the planned security properties and the implemented security mechanisms.

**Design of study** In this part of the evaluation, we focus on the effectiveness of the SecDFD contract verification to answer the following research question.

**Table 2** Results of the mapping after each iteration

| Project | It. | Automated | | Manual | | | |
|---|---|---|---|---|---|---|---|
| | | Precision [%] | Recall [%] | Accept + u | ($\sum$) | Reject | Recall [%] ($\Delta$) |
| jpetstore | 1 | 56.1 | 51.1 | 23 + 3 | (26) | 18 | 57.8 (+6.7) |
| | 2 | 96.4 | 60.0 | 1 + 3 | (30) | 1 | 66.7 (+6.7) |
| | 3 | 96.8 | 66.7 | 0 + 5 | (35) | 1 | 77.8 (+11.1) |
| | 4 | 97.4 | 82.2 | 2 + 3 | (40) | 1 | 88.9 (+6.7) |
| | 5 | 100 | 93.3 | 2 + 3 | (**45**) | 0 | **100** (+6.7) |
| ATM simulation | 1 | 72.0 | 40.0 | 18 + 3 | (21) | 7 | 46.7 (+6.7) |
| | 2 | 67.6 | 51.1 | 2 + 5 | (28) | 11 | 62.2 (+11.1) |
| | 3 | 70.5 | 68.9 | 3 + 5 | (36) | 11 | 80.0 (+11.1) |
| | 4 | 76.6 | 80 | 0 + 4 | (40) | 13 | 88.9 (+8.9) |
| | 5 | 95.5 | 93.3 | 2 + 3 | (**45**) | 2 | **100** (+6.7) |
| Eclipse sec. storage | 1 | 73.0 | 90.5 | 40 + 1 | (41) | 14 | 92.9 (+2.4) |
| | 2 | 67.7 | **100** | 1 + 0 | (**42**) | 12 | – |
| CoCoME | 1 | 27.9 | 77.3 | 17 + 1 | (18) | 44 | 81.8 (+4.5) |
| | 2 | 86.4 | 90.5 | 1 + 1 | (20) | 2 | 90.9 (+0.4) |
| | 3 | 90.9 | 83.3 | 0 + 2 | (**22**) | 4 | **100** (+16.7) |
| iTrust | 1 | 23.5 | 80.0 | 8 + 1 | (9) | 26 | 90.0 (+10.0) |
| | 2 | 81.8 | 90.0 | 0 + 1 | (**10**) | 2 | **100** (+10.0) |

*RQ1. How effective is the proposed approach in the verification of contracts?* It is important to evaluate if the proposed checks can effectively be used in the context of realistic projects. To this aim, we have used open source Java projects, as opposed to illustrative projects. Further, as we are interested in the effectiveness of the proposed compliance checks, we execute the evaluation for all process contracts, encrypt, decrypt, forward, and join. We evaluate the approach with perfectly compliant SecDFDs (i.e., verification results only include convergences, and there are no absence or divergence violations) and with SecDFDs with injected process contracts. In case of the fully compliant SecDFDs, all the detected compliance violations are false positives (FPs). Injecting the process contracts allows us to measure expected compliance violations (e.g., an absence of a join contract), which we mark as true positives (TPs). If the expected compliance violation is not found (according to the injected contract), we mark it as a false negative (FN). Finally, if we find unexpected compliance violations we mark them as false positives (FPs). As a term of measure, we adopt the well-understood precision ($TP/(TP + FP)$) and recall ($TP/(TP + FN)$) of detected compliance violations.

**Execution** As subjects of this evaluation, we use two projects from the introduced test corpus, the Eclipse secure storage and iTrust. The other projects (i.e., jpetstore, ATM simulation, and CoCoME) had less security specifications publicly available. Also, some project implementations did not include any encryption and were less interesting to analyze from the security perspective (e.g., ATM). For both projects, we created one additional SecDFD. In what follows, we refer to the new SecDFDs as Eclipse 2 and iTrust 2. The two SecDFDs created for the study in Sect. 6.1 are Eclipse 1 and iTrust 1. As the created SecDFDs (all four) have been reverse engineered from the implementations, these are perfectly compliant.

First, we apply the contract verification to the two projects. We expect to detect no divergences or absences between the SecDFD and the implementation. Afterward, we inject violations into the systems and check if these are detected. The violations are injected by adding random contracts to the SecDFDs that are not implemented. After every injection, we execute the contract verification and check if the expected violation has been detected, if additional false alarms have been raised, or if expected convergences are not detected any longer. We generate injections of all contract types (encrypt, decrypt, forward, and join). Regardless of the contract type, we inject all possible contracts that have not been specified on the initial SecDFD.

New encrypt and decrypt contracts can be injected independently of each other. An encrypt contract can be injected to every process that has no encrypt contract in the initial SecDFD and a decrypt contract to every process that has no decrypt contract. Accordingly, it can happen that we inject a decrypt contract to a process that has already an encrypt contract and the other way around. For the injection of forward and join contracts, we inject for every process of a SecDFD all possible contracts that are not already specified. To do so, we calculate all possible combinations with one outgoing flow. To calculate the combinations, we consider all incoming and outgoing assets. For instance, for a process with two incoming and two outgoing assets (and no specified forward, or join contract), we inject 6 possible contracts. Every incoming asset can be forwarded to every outgoing asset (4 forward contracts) and the pair of incoming assets can be joined with both outgoing assets as target (2 join contracts). If a combination is equivalent to an existing contract, it is omitted.

**Table 3** Results of evaluating the cryptographic contracts verification

| | Eclipse | | iTrust | | Overall |
|---|---|---|---|---|---|
| | 1 | 2 | 1 | 2 | |
| TPs | 12 | 48 | 59 | 70 | 189 |
| FPs | 0 | 0 | 0 | 0 | 0 |
| FNs | 0 | 0 | 11 | 0 | 11 |
| precision | 100% | 100% | 100% | 100% | **100%** |
| recall | 100% | 100% | 84.28% | 100% | **94.5%** |

**Table 4** Results of evaluating the processing contracts verification

| | Eclipse | | iTrust | | Overall |
|---|---|---|---|---|---|
| | 1 | 2 | 1 | 2 | |
| TPs | 1 | 29 | 55 | 67 | 152 |
| FPs | 0 | 28 | 1 | 10 | 39 |
| FNs | 14 | 29 | 23 | 14 | 80 |
| precision | 100% | 50.88% | 98.21% | 87.01% | **79.58%** |
| recall | 6.67% | 50% | 70.51% | 82.71% | **65.52%** |

**Results** The results of the evaluation are in favor of using our approach to execute security compliance checks between design and implementation.

For the execution of the verification on the fully compliant SecDFDs, we achieved 100% precision and recall. But, the effectiveness of the proposed contracts must also be studied in the context of imperfectly mapped SecDFDs. In what follows, we discuss the effectiveness of the approach in detecting absences of specified contracts. Tables 3 and 4 depict the results of the contract verification based on the injected contracts. We show the results per SecDFD and overall.

For evaluating the verification of encrypt and decrypt contracts, we injected 200 additional encrypt and decrypt contracts into the SecDFDs. Most injected contracts (except 11) were correctly detected as absent. The 11 undetected absent contracts belong to the same SecDFD (of the iTrust project). After investigating them, we noticed that all of them have been injected into processes that already have an encrypt or a decrypt contract. The reason for this defect is that the project-specific specified signature (in the list of well-known cryptographic operations) for encryption is also specified for decryption. As iTrust uses a crypto-function on which a parameter is used for specifying whether a encryption or decryption should be performed, this is a correct classification. Since we only check for at least one method call for encrypt/decrypt, we cannot detect an absence in this particular case.

To evaluate the forward and join checks, we injected 232 contracts (all the possible contract types and combinations for every process) into the SecDFDs. In contrast to the cryptographic contracts verification, the results presented in Table 4 paint a more diverse picture. On the one hand, the processing contracts verification reaches a very good precision (98.21% and 87.01%) and recall (70.51% and 82.71%) on the iTrust project. On the other hand, the verification performs slightly worse when executed on the Eclipse secure storage project. In addition, there is a huge difference between the two SecDFDs on the Eclipse secure storage.

In particular, the verification did not work for the SecDFD shown in Fig. 2 (Eclipse 1). There are two reasons for this poor performance.

First, external entities are not part of the system and cannot be mapped to elements from the system. For example, the external entity *Plugin* in Fig. 2 represents an arbitrary plugin installed into the Eclipse instance that is unknown to the Eclipse secure storage. This arbitrary plugin accesses the secure storage using a Java API specified on implementation level. Similarly, the data can be stored in a cloud, to which access is controlled via an API. In such cases, we attempt at guessing possible incoming flows by considering, e.g., every parameter of the methods mapped to a process as possible source but also all returns of called methods that have not been mapped to any process. For instance, the *Get_value* process (Eclipse 1) is heavily interacting with an external entity and data store which results in very many guesses weakening the results.

Second, despite the reduction when extracting flows (described in Sect. 4.1), the overlapping asset types caused both FPs and FNs. In example, this communication of *Get_value* is implemented by mainly using assets whose mappings are overlapping (mainly strings). In general, representing sensitive objects with string values is prevalent in Eclipse secure storage. This also effected the performance of the processing contracts verification on the second SecDFD (Eclipse 2). Yet, the verification still achieves a recall and precision of 50%. This happened because the asset types of injected contracts overlapped with the asset types of the implemented contracts. For instance, consider two existing and fulfilled forwards of assets that are both mapped to the type *String*. In Fig. 2 for instance, these are the forward of *id* on the *Get_value* process and the forward of the *data* to *encr. data*.[5] In addition to these expected forwards, there are some additional uses of strings that are not representing assets, e.g., a parameter representing a default value in the implementation of the *Get_value* process. Now, we inject a join of *id* and *data* to *encr. data.*. As the default value is a guessed flow, we could easily ignore it before this injection, but now it exactly contributes to the injected join contract and we have to report this contact as convergence. However, we cannot any longer report the forward of *data* as convergence as the flow pattern is now mapped to the injected join contract. Accordingly, we now report a false divergence. In this case, at least the user would have been warned about a violation, but the information about the assets was not entirely correct.

[5] Note that the *Get_value* encrypts the *data* only if it is stored in plain, else it forwards it.

As the iTrust project does not have as many overlapping asset-type mappings and the SecDFDs have less external entities, the results are much better for this subject. Again, the missed violations are mainly due to overlapping asset mappings.

Overall, the contract verification is fairly precise (80%) and reaches the recall of more than 65%. Generally, the contract verification works and is able to bridge the huge gap between early design models and concrete implementations. Though, it suffers from overlapping mappings. Also, missing API specification of the system (i.e., issue of mapping external entities) has a negative impact on the performance of the contract verification.

## 6.3 Evaluation of optimized data flow analysis

The purpose of this study is to evaluate whether using our approach helps to reduce the number of false alarms raised by an existing data flow analyzer.

**Design of study** We investigate the performance of an analysis with FlowDroid [5] initialized with project-specific sources and sinks. To this aim, we built three configurations of sources and sinks. Except in the first configuration (PLAIN), we execute the analyzer *for each SecDFD asset* separately. This experiment was conducted with two projects from Table 1, namely, Eclipse Secure Storage [20] and iTrust [43]. To the best of our knowledge, both projects are free of data flow leaks. Therefore, all the reported leaks by the analyzer are by default labeled as false alarms (FPs). We pose one research question.

*RQ1. To what extent can the mapped design model (with our approach) be used to reduce the number of false alarms raised by a data flow analyzer?*

PLAIN. We execute the analyzer with the list of source signatures shipped with FlowDroid [4,56](herein DEFAULT SOURCES) and sink signatures (herein DEFAULT SINKS). Apart from Java method signatures, this list contains signatures of methods specific to Android source packages. We removed such signatures to avoid unnecessarily searching for them with FlowDroid. Note, that this reduced the list of source signatures from 18,077 to 1,229 and sink signatures from 8,315 to 1,310. As a result of this filtering, the Android SQL database API (SQLite) was also removed. To analyze Java projects, we manually added signatures from the Java SQL API to the above list of sources and sinks.

PARTLY OPT. We execute the analyzer (for each confidential asset) with project-specific source signatures (herein SECDFD SOURCES) and DEFAULT SINKS. The SECDFD SOURCES are extracted per SecDFD asset, as described in Sect. 4. Note that the SECDFD SOURCES are extracted independently, and therefore may not include any of the DEFAULT SOURCES.

FULLY OPT. We execute the analyzer (for each confidential asset) with SECDFD SOURCES and without allowed sink signatures (herein SECDFD SINKS). The list of allowed sink signatures is extracted per SecDFD asset, as described in Sect. 4. The SECDFD SINKS are obtained by *removing* the allowed sink signatures from the DEFAULT SINKS.

The results are compared in only terms of the number of FPs, as no actual leaks (TPs) exist in the analyzed projects. In addition, we measure the number of extracted project-specific source signatures, and the number of removed sink signatures. A false alarm (FP) is a detected leak with a *unique pair of source and sink method signatures*, regardless of the access path where the leak is detected. The rationale for counting unique signature pairs is that comparing access paths would be computationally expensive and not useful for the purpose of this study. For instance, consider an implementation of a function where the number of recursive calls depends on a conditional. In this case, at least two access paths (when the conditional evaluates to `true` and `false`) are detected. But the DFD does not specify such level of detail, thus we cannot distinguish between the access paths of the detected data leaks. The false alarms are aggregated per SecDFD, to enable comparison with the PLAIN configuration.

As we execute the analysis for each SecDFD asset, we measure the project specific sources and sinks in the same manner. Specifically, to measure the number of project-specific sources we count each discovered source signature per SecDFD asset. Similarly, to observe the number of times we are able to remove an allowed sink, we count each signature which has been removed for a unique asset.

**Listing 1** Configuration of FlowDroid used in this study

```
Infoflow result = new Infoflow("", false, null)
    ;
result.setSootConfig((options, config) -> {
config.setCallgraphAlgorithm(CallgraphAlgorithm
    .AutomaticSelection);
config.setImplicitFlowMode(ImplicitFlowMode.
    AllImplicitFlows);
config.setAliasingAlgorithm(AliasingAlgorithm.
    FlowSensitive);
config.setStopAfterFirstKFlows(100);
});
result.setTaintWrapper(new EasyTaintWrapper(
    Collections.emptyMap()));
return result;
```

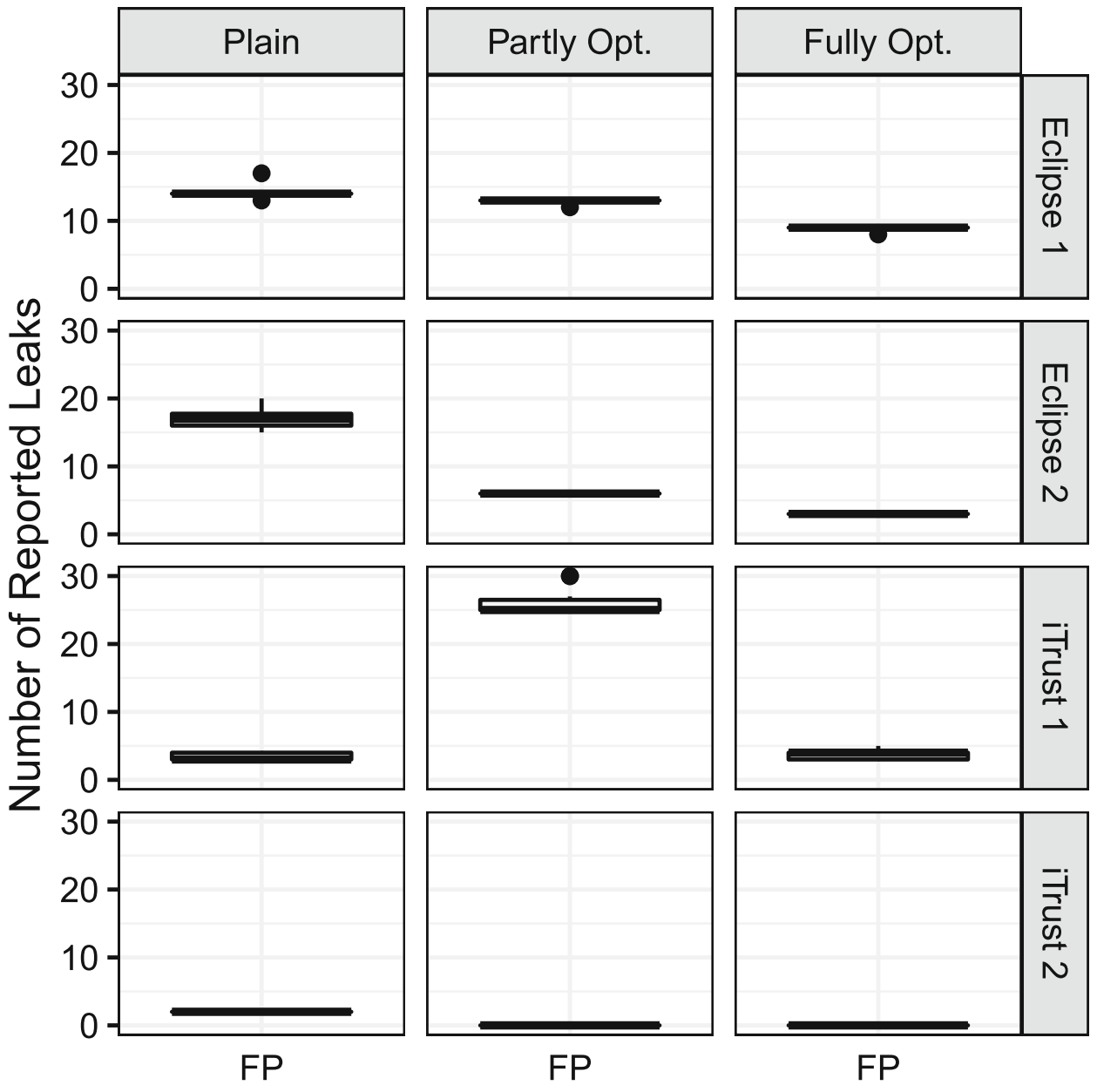

**Fig. 8** False alarms (FPs) raised by the analyzer after three configurations of sources and sinks per SecDFD (Eclipse Secure Storage on top, iTrust on bottom)

**Table 5** Average false alarm reduction for the different configurations (aggregated per project)

| Configuration | FPs on Eclipse | | FPs on iTrust | | Overall |
|---|---|---|---|---|---|
| Plain | 15.65 | | 2.7 | | 9.18 |
| Partly Opt. | 9.45 | (↓ 60%) | 13.1 | (↑ 485%) | 11.28 |
| Fully Opt. | 5.95 | (↓ 37%) | 1.9 | (↓ 85%) | 3.93 |
| Total | | (↓ **62**%) | | (↓ 30%) | (↓ **57**%) |

**Execution** Both projects used in this study include two SecDFDs, representing two different scenarios. Listing 1 shows how we configured FlowDroid for all our executions. This configuration was set up to achieve the best performance and most conservative analysis, in accordance with the literature [3]. We configure FlowDroid to use the default call-graph construction algorithm (SPARK). In addition, we have enabled implicit flow tracking and flow-sensitive aliasing. Note that, without tracking implicit flows, FULLY OPT. produces no false alarms, while PLAIN still reports many. Finally, we limit the static analysis to the projects, excluding third-party libraries (line 11 in Listing 1), and stop the analyzer after identifying 100 leaks per run. We have implemented and executed the experiments using the JUnit Plugin Test framework with a limit of 6 GB of memory consumption (for each execution of the analyzer). The amount of allowed memory and the maximum number of identified leaks were determined empirically. We have executed random parts of the experiment with different configurations repeatedly and did not get different results.

**Results** Figure 8 shows the false alarms raised by the analyzer after three configurations per SecDFD model. The average number of false alarms is aggregated per project in Table 5, and the change in the number of false alarms is presented. The main takeaway of the evaluation is that using our approach we were able to (a) extract project-specific sources of secret data, and (b) reduce the number of false alarms (up to 62%) raised by the data flow analyzer. First, we discuss the reduction with only project-specific sources. Second, we discuss the reduction with removing allowed sinks.

*RQ1.* Our measurements from the PARTLY OPT. configuration show that deriving project-specific sources from the SecDFD is possible and can reduce the number of FPs. For instance, in case of Secure Storage we achieved an average 60% reduction in false alarms (Table 5). However, adding project-specific sources can also lead to a rise in false alarms (as observed on iTrust). The number of project-specific sources is realistic considering the project size (i.e., 11 for Secure Storage and 10 for iTrust). In addition, the project-specific source methods are in fact accessing sensitive resources (e.g., the `org.eclipse.equinox.internal.security.storage. SecurePreferences. get (String, String, SecurePreference Container):String` is called when fetching the cashed confidential credentials.) But, the derived sources depend heavily on the mappings. Since iTrust is implemented with the dynamic Java Server Pages, FlowDroid cannot analyze the entire behavior of the program. Therefore, we are only able to reduce the number of FPs after removing allowed sinks.

We found that the number of FPs can be further reduced by removing allowed sinks from the list of sinks passed to the analyzer (FULLY OPT. configuration). We have been able to remove 3 sinks (all from `java.lang` package) for Eclipse Secure Storage and 36 sinks (all from `java.sql` package) for the iTrust project. These sinks were included in the previous configurations, but were derived in this configuration as allowed for certain SecDFD assets. In particular, we observed a further 37% average reduction in FPs for the Eclipse Secure Storage project, when comparing the analysis results to the previous configuration (PARTLY OPT.). Compared to the first configuration (PLAIN), considering only project-specific sources and removing allowed sinks reduced the number of false alarms on average by 62%. As project-specific sources were hard to find for the iTrust project, we compare the analysis results to the initial configuration (PLAIN). Removing the allowed sinks in iTrust reduced the number of FPs on average by 30%.

## 7 Discussion and threats to validity

This section discusses the results and examines the threats to validity of the proposed approach.

### 7.1 Discussion

**Generalizability** In the domain of safety critical systems, reliability requirements are highly prioritized and organizations have mature compliance processes in place. In contrast to security, reliability engineers analyze how the system behaves in case of failures. However, the semi-automated mappings (and the accompanying tool support) are general enough and could be used as a foundation to develop compliance checking for *reliability and safety properties*. For example, the implementation of processes with outgoing data flows can be checked to never return data with null fields. In addition, the approach could be extended to support compliance analyses of real-time properties of distributed systems, such as *liveliness* (the implementation could be checked for unbound loops). Similarly, compliance to *performance* requirements could be checked by tracing the implemented information flows and analyzing their distribution. Though, besides the application code, additional sources of information about the deployment (e.g., network policies) and runtime information may be required for comparing the intended levels of performance to actual computing load.

The security analysis (Sect. 4) could also be extended to check compliance of *integrity* as a dual to confidentiality [10] (instead of confidential sources, the developer needs to specify the trusted sources and sinks). Further, our approach can be extended to support compliance analysis of *accountability* requirements of asset manipulations (as stipulated by the design) with respect to the implemented logging mechanisms.

**Algorithm complexity** Worse case complexity of the heuristic algorithm for finding the mappings is exponential; however, the SecDFD models are much smaller compared to the program model. Also, the rules for corresponding elements (Sect. 3.1) significantly reduce the search complexity. In contrast to the implementation of forward and join checks, the complexity of checking an encrypt/decrypt contract is linear. Though a complete complexity analysis of the forward and join checks was not conducted, we observed that the number of types, members, and accesses are the most influential factors. The reader should note that we do not attempt to trace the entire implementation, but only targets of the mappings (smaller chunks); therefore, the runtime complexity of the checks is fair, also for realistic-sized applications.

**Scalability** Our evaluation of the mappings suggests no major scalability issues when applying the approach to realistic applications. The user seldom defined mappings manually (on average, 75% of mappings were suggested by the algorithm) and in most applications, accepted more mappings than rejected. In principle, the approach could still be used with some unaccepted mappings. Not accepting correct mappings would prevent the algorithm in further exploring the search space, but the exact impact on the precision and recall must be further investigated.

### 7.2 Threats to validity

The main threat to *external validity* is our selection of samples, based on a limited number of open source projects, partially originating from a teaching context. Regarding the validity of the studies conducted to evaluate the security compliance checks, the open source projects do not contain well-known data flow leaks, thus we consider them secure. The rationale for our selection was the manual effort that was required for creating the ground truth of our technique, a full mapping between high-level DFD elements and low-level program elements. However, as a result, the generalizability of the results to larger project in other domains is limited. To mitigate this threat, the considered projects were chosen to be representative for realistic projects by providing a good documentation, including architectural information (such as, wikis, use cases, scenarios, requirements, state charts, and the like). The available documentation enabled building good design models, close to the intended architecture. Further, we partly mitigate this threat by experimenting with contract injections as part of our evaluation. We plan to extend the evaluation in the future to include a more comprehensive set of projects.

Regarding *internal validity*, the main threat of our evaluation is researcher bias. In the absence of pre-existing ground truths and design models, the ground truth and design models for our evaluation were created manually by the authors, possibly introducing a risk of creating a biased result. To mitigate this threat, the ground truths and the design-level models were carefully discussed between all authors. The created models and ground truths are of similar size and complexity and are available online [53].

With respect to *construct validity*, we consider the threat of misinterpreting divergence, absence, and convergence compliance violations in the context of design-level models, implementation-level models, and violations detected by static code analysis. However, to the best of our knowledge, our interpretations are in-line with the existing literature [15].

## 8 Related work

First, we discuss two most related works with respect to security compliance of DFDs, and leveraging specifications to optimize data flow analysis. Similar to our work, these approaches are difficult to classify as forward or reverse engineering solutions. Next, we position our work in the context of *forward* and *reverse engineering* literature.

More than a decade ago, Abi-Antoun et al. [2] proposed conformance checks between the implementation and DFDs.

The authors automatically extract a DFD (i.e., the *source DFD*) from the implementation. Next, the user specifies a mapping (using Reflexion Models) between a manually created *high-level DFD* and the source DFD, which is then used to uncover inconsistencies. The notion of extracting the source DFD is similar to our extraction of the implemented data flows. In contrast to the mappings with Reflexion Models, our mappings are semi-automated using heuristics. Further, the security analysis in [2] is performed on the level of the DFD, while our security compliance checks are developed by means of static code analysis. To the best of our knowledge, this work is the sole attempt at implementing security compliance checks between the SecDFD and its implementation.

Recently, static code analysis techniques have been developed to assure GDPR compliance of the implemented systems with respect to privacy specifications [24,26,31]. Most relevant to our work is the proposed approach by Ferrara et al. [26] which uses the privacy policy to fine tune and execute a taint analysis. The authors evaluate the approach by executing a prototype analysis on a benchmark application. Deriving the sources and sinks from the privacy policy is similar to our idea of maintaining allowed sinks for each SecDFD asset. But, the required GDPR policy needs to be specified on the level of implementation (e.g., concrete fields as sources, and API method signatures for sinks). In contrast, our approach can derive project-specific sources and allowed sinks from the design and also performs security compliance checks with respect to the design model.

UML models have been extensively studied in the context of **forward engineering** solutions for checking security compliance.

Muntean et al. [44] extend the UML statecharts with security annotations (such as source function, sink function, declassified parameter, etc.), generate the source code in C, and implement static checks (using the Smtcodan engine) to detect data flow violations. Similar to our work, the authors leverage security information from the design to execute a static analysis, and lift the detected violations back to the user (they display them with sequence diagrams). However, compared to DFDs, the gap between statecharts and source code is smaller (e.g., DFDs can not express conditional data flows, or sequence of data flows). Further, our approach with correspondence mappings can be used on existing projects (no code generation is necessary).

IFlow [35] is an approach for specifying and analyzing information flow properties in distributed Java applications. The proposed approach extends the UML model with information flow properties and uses it to generate a Java code skeleton, and transform it to a formal model supporting an interactive theorem prover. The Java code skeleton (and manually completed program) can be checked for standard information flow properties, such as non-interference, using an existing framework (i.e., JOANA). Similar to this work, IFlow requires the developer to provide the security information in the model and leverages an existing static analyzer. But, IFlow is model-driven and analyzes non-interference in a more formal setting.

Fourneret et al. [27] combine model-based security analyses using UMLsec [34] with the generation of security tests. Security properties are specified and verified on UML state machines. These models are then used to generate tests for the implemented system. In contrast to us, the considered state machines have to be very close to the implementation. Further, Ramadan et al. [55] use model transformation to automatically generate security-annotated UML class models [34] from security-annotated BPMN models.

For the classical **reverse engineering** scenario from source code to UML class models, Peldszus et al. [51] propagate hand-crafted security annotations from source code to the corresponding elements in automatically extracted class models.

Scoria [69] is a semi-automated approach for extracting and analyzing the Owner Object Graph annotated with security properties (i.e., SecGraph) to find security flaws in the architecture. First, The SecGraph is extracted from a manually annotated implementation. Second, software architects can optionally refine the SecGraph with additional annotations. Finally, software architects can design queries to analyze the SecGraph. Similar to our work, Scoria is an iterative semi-automated approach analyzing security on abstracted code representation. However, our work does not rely on code annotations and executes the security compliance checks by means of static analysis.

Jasser [32] recently proposed an approach for analyzing system behavior and detecting its discordance with a set of useful security rules. The security rules (modeled as Linear Time Logic (LTL) properties) are expressed with a controlled natural language for describing architectural constraints. The system behavior is extracted by means of dynamic analysis, using aspect-oriented programming. Finally, before the security rules can be executed, the source-level elements are manually mapped to the architectural elements. On a high-level, the idea of our work relates Jassers approach, in that, an abstracted representation of code is mapped to a higher-level model to analyze security compliance. In comparison, our approach supports an automated discovery of such mapping and studies the compliance of *static* security properties in the implementation.

Manual security reviews can be aided by automated static (or hybrid) program analysis. Static Application Security Testing (SAST) [25] tools aim to analyze the program code of a software component and automatically report the violations to developers, removing the need for security experts reviewing large code bases. Our approach relates to such mechanisms in that it leverages static code analysis to eval-uate security of an implemented system. But, the SAST analyzes security of the implementation, while our approach focuses on analyzing the compliance of implemented security to the intended (designed) security. Further, SAST tools need to still be configured by security experts, whereas our approach automatically derives project-specific sources and sinks from the SecDFD model.

Duarte et al. [19] propose to use context information of execution sequences for the extraction of labeled transition system models from source code. While the authors motivate their approach with the need for correspondence between models and code, they only discuss the possibility to analyze the models using existing tooling. The compliance checks introduced by Duarte et al. [19] are performed similarly to the checks developed in this work. In contrast, our approach supports compliance checks between models and code. Regarding the preparation for compliance checks, Duarte et al. reverse engineer models that can be checked or compared to existing models. In contrast, we recreate a mapping between existing models and their implementation. This already includes a comparison with the existing models.

Beyond the security scope of this work, conformance checking is generally a well-studied topic in model-driven engineering. Paige et al. [46] use meta-models as the common reference point to enable conformance checks between diagrams representing different views on a system. Diskin et al. [17] present a framework for global consistency checks of heterogeneous models based on constraints. By supporting the explicit specification of overlaps between the considered models, they avoid the need for a global meta-model. Expanding on this work, König and Diskin [37] improve the efficiency of this approach by supporting an early localization of relevant parts of the models whose consistency is to be checked. Reder and Egyed [57] propose an efficient approach to consistency checking based on predefined consistency rules. Estanol et al. [22] developed an approach to check the conformance of process implementation to UML and OCL models by translating them into petri-nets, and executing existing conformance checking techniques. However, none of these works address security compliance checking between design and its implementation.

## 9 Conclusion and future work

This work has introduced a novel approach for tackling the problem of automating the code-level verification of planned security mechanisms. In particular, we have developed a solution with tool support for executing security compliance checks between an abstract design model (the SecDFD) and its implementation (in Java). To this aim, we developed a user-in-the-loop approach for finding corresponding elements based on heuristically computed suggestions. Once defined, the correspondence mappings are leveraged for an automated security analysis of the implementation against the design. First, two types of security compliance checks are executed: a rule-based check for a set of cryptographic operations, and a local data flow check for data processing contracts specified in the model. Second, the mapped design is leveraged to initialize and execute a state-of-the-art data flow analyzer over the entire Java project. The results of the compliance checks (convergence, absence, and divergence) are lifted to the attention of the user via the user interface of our tool.

Our approach was evaluated with three experiments on open source Java projects (five in the first experiment and two in the second and third), focused on assessing the performance from different angles. First, our evaluation has shown a high precision (87.2%) of the automated suggestions of mappings. Second, the rule-based security compliance checks are very precise (100%) and rarely overlook implemented cryptographic operations (recall is 94.5%). In addition, the local data flow checks are fairly precise (79.6%), but may overlook some implemented flows (recall is 65.6%), due to the large gap between the design and implementation. Finally, our approach enables a project-specific data flow analysis with up to 62% less false alarms.

Regarding future improvements, we note that extending the SecDFD with strongly typed assets could improve the performance of the security compliance checks. Strongly typed SecDFD assets could be mapped to the implementation more precisely, which would make the local data flow checks cleaner. In addition, the missing mappings to the external entities could be better approximated by relying on parsed API specifications (e.g, JavaDoc). Finally, the evaluation of the security checks could be improved by including more open source projects, especially projects with well-known data leaks.

**Acknowledgements** The work presented in this article is part of the Ph.D. theses of Katja Tuma [65] and Sven Peldszus [47]. This research was partially supported by Deutscher Akademischer Austauschdienst (DAAD), the H2020- projects AssureMOSS (grant agreement No 952647), TRUSTS and Qu4lity that received funding from the European Union's Horizon 2020 research and innovation programme, and the BMWi-project IIP Ecosphere. This paper reflects only the author's view and the Commission is not responsible for any use that may be made of the information contained therein.

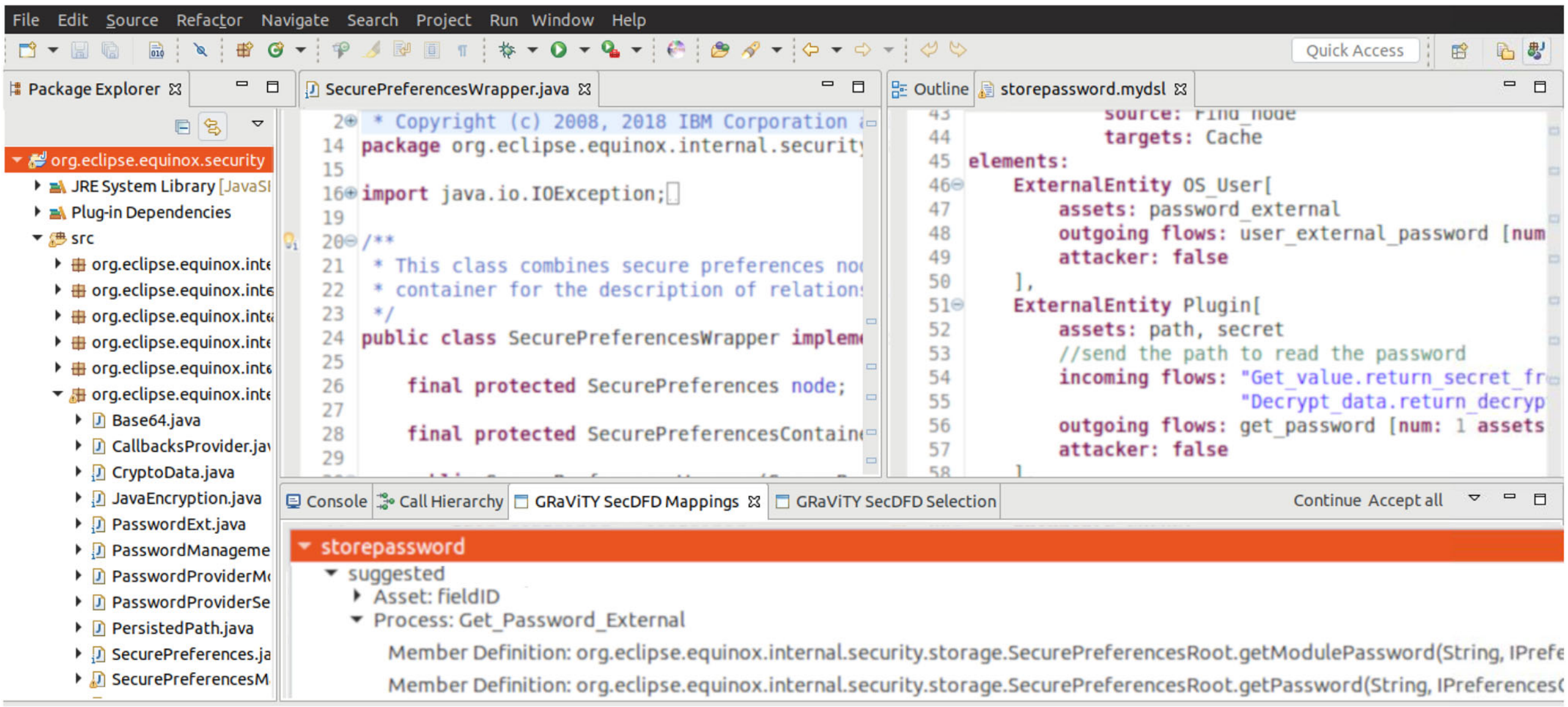

**(a)** UI with the textual syntax

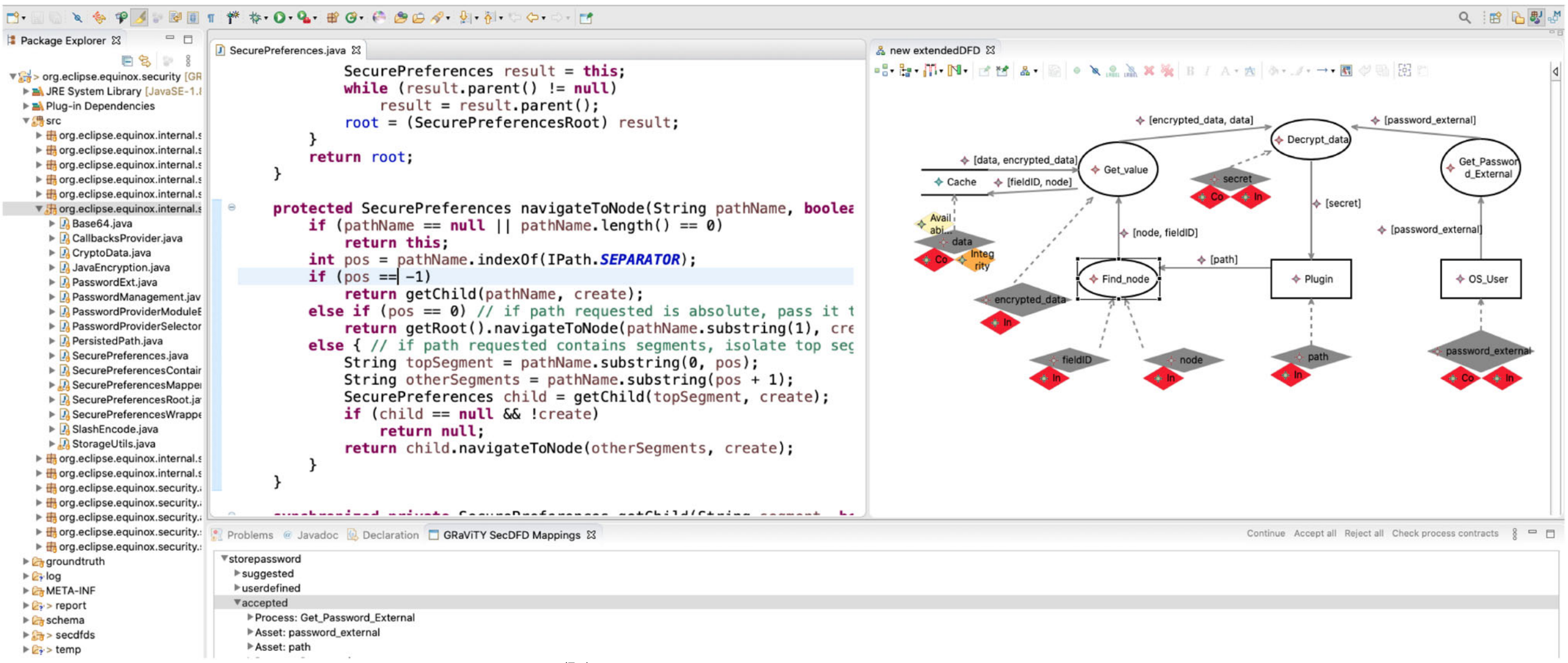

**(b)** UI with the graphical syntax

**Fig. 9** Screenshots of the UI in Eclipse

## A Using the tool

The target audience of the tool are software developers with training in the principles of software architecture. After the installation of the required packages, the program is started as a running Eclipse instance. First, the developers import the desired Java project. Second, they manually create one or several SecDFDs for representing the high-level architecture and security properties of the project. They can do so with a textual or graphical syntax (one can be generated from the other). Figure 9 shows screenshots of the user interface after this step is completed. On the left hand side of the figures, users can see the Package Explorer. The top two windows are used for displaying the source code (left) and the SecDFD (right). The bottom windows are used for displaying and defining the mappings. Next, using context menu entries, the developers trigger the automated generation of a PM from the source code, and start the first iteration of the semi-automated process for mapping the SecDFD elements to source code elements (see Sect. 3).

At the start of each iteration, the developers are shown a list of suggested mappings. Since one DFD element is usually mapped to several program elements, the results are grouped by the DFD elements. For each DFD element, the list of mapped PM elements is shown, each with its path in the source code. The developers can interact with the tool by accepting, rejecting, and manually defining mappings. A suggested mapping is accepted or rejected with a right-click on the entry and selecting *accept* or *reject*, respectively. Once a mapping is accepted, corresponding in-line markers are created on the SecDFD and in the source code. Double-clicking a mapping will open the correct source file and navigate to the correct line in the file. Accepted mappings can always be rejected. If all the suggested mappings are correct, the developers can select *accept all*. Rejected mappings will never be suggested again. Manual definition works by right-clicking and selecting *Map Selection to SecDFD* on source code elements. At the end of the iteration, developers can either stop or select *continue* to trigger a new search refining the present mapping.

Finally, the developers can execute security compliance checks by pressing a button. The contract violations and leaks identified by FlowDroid are presented to the user with error and warning markers on the SecDFD model. At any moment, the developers can extend the list of project-specific methods signatures for cryptographic operations, and execute the checks again. Similar to manual definition, they can right-click the source code elements and select the appropriate menu item.

## References

1. Abe, T., Hayashi, S., Saeki, M.: Modeling security threat patterns to derive negative scenarios. In: APSEC, pp. 58–66 (2013)
2. Abi-Antoun, M., Wang, D., Torr, P.: Checking threat modeling data flow diagrams for implementation conformance and security. In: ASE, pp. 393–396 (2007)
3. Arzt, S.: Static data flow analysis for android applications. Ph.D. thesis, Technische Universität Darmstadt (2017)
4. Arzt, S., Rasthofer, S., Bodden, E.: SuSi: a tool for the fully automated classification and categorization of android sources and sinks. Tech. Rep. TUDCS-2013-0114, University of Darmstadt (2013)
5. Arzt, S., Rasthofer, S., Fritz, C., Bodden, E., Bartel, A., Klein, J., Le Traon, Y., Octeau, D., McDaniel, P.: Flowdroid: precise context, flow, field, object-sensitive and lifecycle-aware taint analysis for android apps. ACM Sigplan Notices **49**(6), 259–269 (2014)
6. Axway Software, BizAgi Ltd, Bruce Silver Associates, IDS Scheer, International Business Machinesand MEGA International, Model Driven Solutions, Object Management Group, Oracle, SAP AG, Software AG Inc, TIBCO, Unisys (2014) Business Process Model And Notation (BPMN). OMG Standard formal/13-12-09, Object Management Group (OMG), version 2.0.2
7. Baca, D., Petersen, K., Carlsson, B., Lundberg, L.: Static code analysis to detect software security vulnerabilities-does experience matter? In: ARES, pp. 804–810. IEEE (2009)
8. Berger, B.J., Sohr, K., Koschke, R.: Extracting and analyzing the implemented security architecture of business applications. In: CSMR, pp. 285–294 (2013)
9. Bernsmed, K., Jaatun, M.G.: Threat modelling and agile software development: identified practice in four Norwegian Organisations. In: Cyber Security, pp. 1–8. IEEE (2019)
10. Biba, K.J.: Integrity considerations for secure computer systems. Tech. rep., MITRE CORP, Bedford, MA (1977)
11. Bjork, R.C.: ATMExample. http://www.math-cs.gordon.edu/local/courses/cs211/ATMExample/ (2020)
12. Bodden, E.: Inter-procedural Data-flow Analysis with ifds/ide and Soot. In: SOAP, pp. 3–8 (2012)
13. Bürger, J., Strüber, D., Gärtner, S., Ruhroth, T., Jürjens, J., Schneider, K.: A framework for semi-automated co-evolution of security knowledge and system models. JSS **139**, 142–160 (2018)
14. Cook, S., Bock, C., Rivett, P., Rutt, T., Seidewitz, E., Selic, B., Tolbert, D.: UML Superstructure Specification. OMG Standard formal/2017-12-05, Object Management Group (OMG), version 2.5.1 (2017)
15. De Silva, L., Balasubramaniam, D.: Controlling software architecture erosion: a survey. JSS **85**(1), 132–151 (2012)
16. Deng, M., Wuyts, K., Scandariato, R., Preneel, B., Joosen, W.: A privacy threat analysis framework: supporting the elicitation and fulfillment of privacy requirements. RE **16**(1), 3–32 (2011)
17. Diskin, Z., Xiong, Y., Czarnecki, K.: Specifying overlaps of heterogeneous models for global consistency checking. In: Models, pp. 165–179 (2010)
18. Dougherty, C., Sayre, K., Seacord, R.C., Svoboda, D., Togashi, K.: Secure design patterns. Tech. rep., Carnegie-Mellon University Pittsburgh, Software Engineering Institute (2009)
19. Duarte, L.M., Kramer, J., Uchitel, S.: Using contexts to extract models from code. SoSyM **16**, 523–557 (2017)
20. Eclipse Contributors. Eclipse Documentation—Secure Storage. https://help.eclipse.org/2020-06/topic/org.eclipse.platform.doc.user/reference/ref-securestorage-start.htm (2020)
21. Ehrig, H., Rozenberg, G., Kreowski, H.J.: Handbook of Graph Grammars and Computing by Graph Transformation, vol. 3. World Scientific, Singapore (1999)
22. Estañol, M., Munoz-Gama, J., Carmona, J., Teniente, E.: Conformance checking in uml artifact-centric business process models. SoSyM **18**(4), 2531–2555 (2019)
23. Faily, S., Scandariato, R., Shostack, A., Sion, L., Ki-Aries, D.: Contextualisation of data flow diagrams for security analysis. In: GraMSec, pp. 186–197 (2020)
24. Fan, M., Yu, L., Chen, S., Zhou, H., Luo, X., Li, S., Liu, Y., Liu, J., Liu, T.: An empirical evaluation of GDPR compliance violations in android mHealth apps. In: ISSRE, pp. 253–264 (2020)
25. Felderer, M., Büchler, M., Johns, M., Brucker, A.D., Breu, R., Pretschner, A.: Security testing: a survey. In: Advances in Computers, vol. 101, pp 1–51. Elsevier (2016)
26. Ferrara, P., Olivieri, L., Spoto, F.: Tailoring taint analysis to GDPR. In: APF, pp. 63–76. Springer (2018)
27. Fourneret, E., Ochoa, M., Bouquet, F., Botella, J., Jurjens, J., Yousefi, P.: Model-based security verification and testing for smart-cards. In: ARES, pp. 272–279 (2011)
28. Goseva-Popstojanova, K., Perhinschi, A.: On the capability of static code analysis to detect security vulnerabilities. IST **68**, 18–33 (2015)
29. Hebig, R., Quang, T.H., Chaudron, M.R., Robles, G., Fernandez, M.A.: The quest for open source projects that use UML: mining GitHub. In: Models, pp. 173–183 (2016)
30. Heinrich, R., Rostami, K., Reussner, R.: The Cocome platform for collaborative empirical research on information system evolution. Tech. Rep. 2016,2, Karlsruhe Institute of Technology (2016)
31. Hjerppe, K., Ruohonen, J., Leppänen, V.: Annotation-based static analysis for personal data protection. In: IFIP, pp. 343–358. Springer (2019)
32. Jasser, S.: Enforcing architectural security decisions. In: ICSA, pp 35–45. IEEE (2020)

33. Jung, R., Heinrich, R., Taspolatoglu, E., Pöppke, T.: CoCoME. https://github.com/cocome-community-case-study (2020)
34. Jürjens, J.: Secure Systems Development with UML. Springer, Berlin (2005)
35. Katkalov, K., Stenzel, K., Borek, M., Reif, W.: Model-driven development of information flow-secure systems with IFlow. In: SocialCom, pp. 51–56. IEEE (2013)
36. Klieber, W., Flynn, L., Bhosale, A., Jia, L., Bauer, L.: Android taint flow analysis for app sets. In: SOAP, pp. 1–6 (2014)
37. König, H., Diskin, Z.: Efficient consistency checking of interrelated models. In: ECMFA, pp. 161–178 (2017)
38. Levenshtein, V.I.: Binary codes capable of correcting deletions, insertions, and reversals. Sov. Phys. Dokl. **10**(8), 707–710 (1966)
39. Li, L., Bissyandé, T.F., Papadakis, M., Rasthofer, S., Bartel, A., Octeau, D., Klein, J., Traon, L.: Static analysis of android apps: a systematic literature review. IST **88**, 67–95 (2017)
40. Lund, M.S., Solhaug, B., Stølen, K.: Model-Driven Risk Analysis: The Coras Approach. Springer, Berlin (2011)
41. Macher, G., Armengaud, E., Brenner, E., Kreiner, C.: A review of threat analysis and risk assessment methods in the automotive context. In: SAFECOMP, pp. 130–141 (2016)
42. Mebus, D.: Objektorientierte high-level Datenflussanalyse. Master's thesis, University of Koblenz-Landau (2019)
43. Meneely, A., Smith, B., Williams, L.: iTrust electronic health care system case study. https://github.com/ncsu-csc326/iTrust (2020)
44. Muntean, P., Rabbi, A., Ibing, A., Eckert, C.: Automated detection of information flow vulnerabilities in UML state charts and C code. In: QRS-C, pp. 128–137. IEEE (2015)
45. MyBatis. JPetStore. http://www.mybatis.org/jpetstore-6/ (2020)
46. Paige, R.F., Brooke, P.J., Ostroff, J.S.: Metamodel-based model conformance and multiview consistency checking. TOSEM **16**(3), 11 (2007)
47. Peldszus S (2021) Security compliance in model driven development of software systems in presence of long-term evolution and variants. PhD thesis, University of Koblenz-Landau
48. Peldszus, S., Kulcsár, G., Lochau, M.: A solution to the Java refactoring case study using eMoflon. In: TTC, pp. 118–122 (2015)
49. Peldszus, S., Kulcsár, G., Lochau, M., Schulze, S.: Incremental co-evolution of Java programs based on bidirectional graph transformation. In: PPPJ, pp. 138–151 (2015)
50. Peldszus, S., Kulcsár, G., Lochau, M., Schulze, S.: Continuous detection of design flaws in evolving object-oriented programs using incremental multi-pattern matching. In: ASE (2016)
51. Peldszus, S., Strüber, D., Jürjens, J.: Model-based security analysis of feature-oriented software product lines. In: GPCE (2018)
52. Peldszus, S., Tuma, K., Strüber, D., Jürjens, J., Scandariato, R.: Secure data-flow compliance checks between models and code based on automated mappings. In: Models, pp. 23–33. IEEE (2019)
53. Peldszus, S., Tuma, K., Strüber, D., Scandariato, R., Jürjens, J.: Implementation and evaluation data. https://github.com/SvenPeldszus/GRaViTY-SecDFD-Mapping (2020)
54. Peldszus, S., et al.: GRaViTY program model. http://gravity-tool.org (2020)
55. Ramadan, Q., Salnitri, M., Strüber, D., Jürjens, J., Giorgini, P.: From secure business process modeling to design-level security verification. In: Models, pp. 123–133 (2017)
56. Rasthofer, S., Arzt, S., Bodden, E.: A Machine-learning approach for classifying and categorizing android sources and sinks. In: NDSS Symposium (2014)
57. Reder, A., Egyed, A.: Incremental consistency checking for complex design rules and larger model changes. In: Models, pp. 202–218 (2012)
58. Ruland, S., Kulcsár, G., Leblebici, E., Peldszus, S., Lochau, M.: Controlling the attack surface of object-oriented refactorings. In: FASE, pp. 38–55 (2018)
59. Saini, V., Duan, Q., Paruchuri, V.: Threat modeling using attack trees. CCSC **23**(4), 124–131 (2008)
60. Santos, J.C.S., Tarrit, K., Mirakhorli, M.: A catalog of security architecture weaknesses. In: Proceedings of the International Conference on Software Architecture Workshops (ICSAW), pp. 220–223. IEEE Computer Society (2017). https://doi.org/10.1109/ICSAW.2017.25
61. Scandariato, R., Wuyts, K., Joosen, W.: A descriptive study of Microsoft's threat modeling technique. RE **20**(2), 163–180 (2015)
62. Shostack, A.: Threat Modeling: Designing for Security. Wiley, Hoboken (2014)
63. Sion, L., Yskout, K., Van Landuyt, D., Joosen, W.: Solution-aware data flow diagrams for security threat modeling. In: SAC, pp. 1425–1432 (2018)
64. Tuma, K., Scandariato, R.: Two architectural threat analysis techniques compared. In: ECSA, pp. 347–363 (2018)
65. Tuma K (2021) Efficiency and Automation in Threat Analysis of Software Systems. PhD thesis, Chalmers University of Technology and Gothenburg University
66. Tuma, K., Calikli, G., Scandariato, R.: Threat analysis of software systems: A systematic literature review. JSS **144**, 275–294 (2018)
67. Tuma, K., Balliu, M., Scandariato, R.: Flaws in flows: unveiling design flaws via information flow analysis. In: ICSA, pp. 191–200 (2019)
68. Vallee-Rai, R., Hendren, L.J.: Jimple: simplifying Java bytecode for analyses and transformations. Tech. rep., McGill University (1998)
69. Vanciu, R., Abi-Antoun, M.: Finding architectural flaws using constraints. In: ASE, pp. 334–344. IEEE (2013)
70. Wolf, T., Dahyabhai, N., Sohn, M., et al.: EGit—user guide. https://wiki.eclipse.org/EGit/User_Guide (2019)

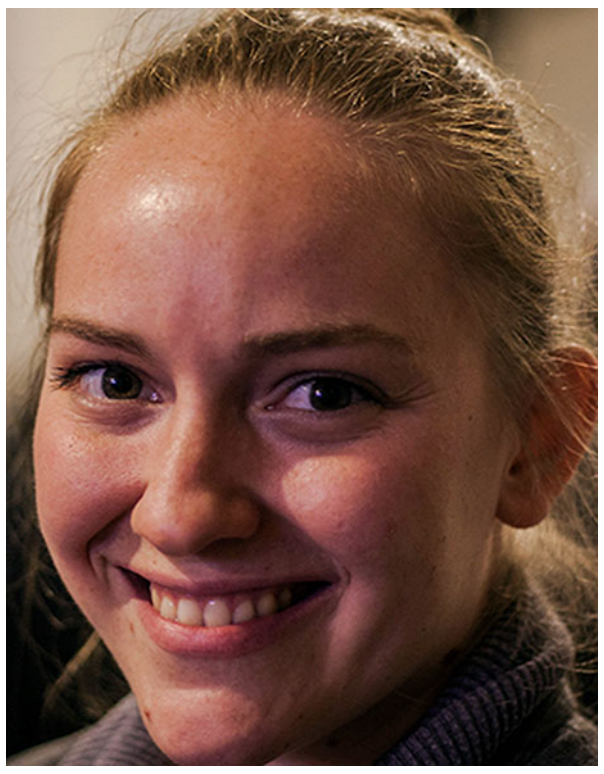

**Katja Tuma** is an assistant professor at the Vrije University in Amsterdam. She obtained her Ph.D. in Computer Science and Engineering from the University of Gothenburg in Sweden. Katja has collaborated with several research groups (e.g., at KTH, KU Leuven, and University of Koblenz-Landau) and has worked closely with the automotive industry. She takes an active role in the community, serving as PC member and reviewing for top scientific venues. Her research is at the intersection of software engineering, security, and risk analysis, with a particular interest in experimentation with threat and risk analysis, analysis automation and compliance to intended security.

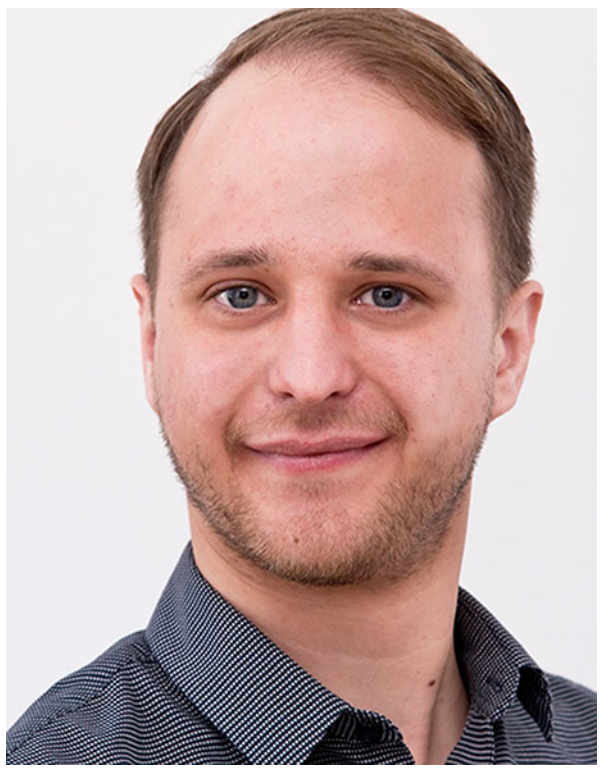

**Sven Peldszus** is a postdoctoral researcher at Ruhr University Bochum and worked before at the Institute for Software Technology at the University of Koblenz-Landau. He received a master's degree from the University of Darmstadt and submitted his PhD thesis at University of Koblenz-Landau. His research interests include the continuous tracing of non-functional requirements throughout the software life cycle and the model-based security and software quality analysis in variant-rich systems.

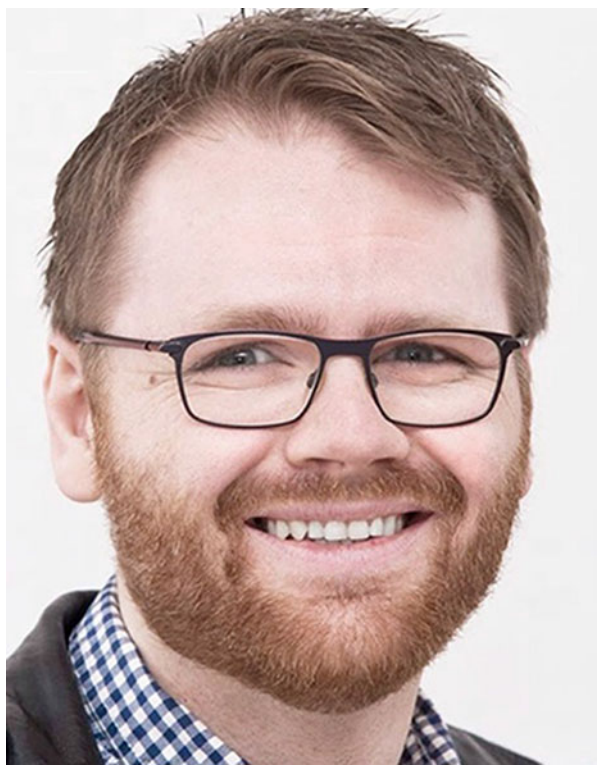

**Daniel Strüber** is a senior lecturer at the Computer Science and Engineering department of Chalmers and Gothenburg University, Sweden. He is also affiliated with Radboud University in Nijmegen, the Netherlands. His research interests are in model-driven engineering, AI engineering, and variant-rich systems. He was awarded his doctoral degree from Philipps University Marburg, Germany, and worked as a postdoctoral researcher at University of Koblenz and Landau, Germany, and Gothenburg University, Sweden. He is a co-author of over 75 papers with six Best Paper Awards. He has been a Program Committee member of several leading conferences, including FASE, MODELS, and SPLC. He is the lead developer of Henshin, an internationally used model transformation language.

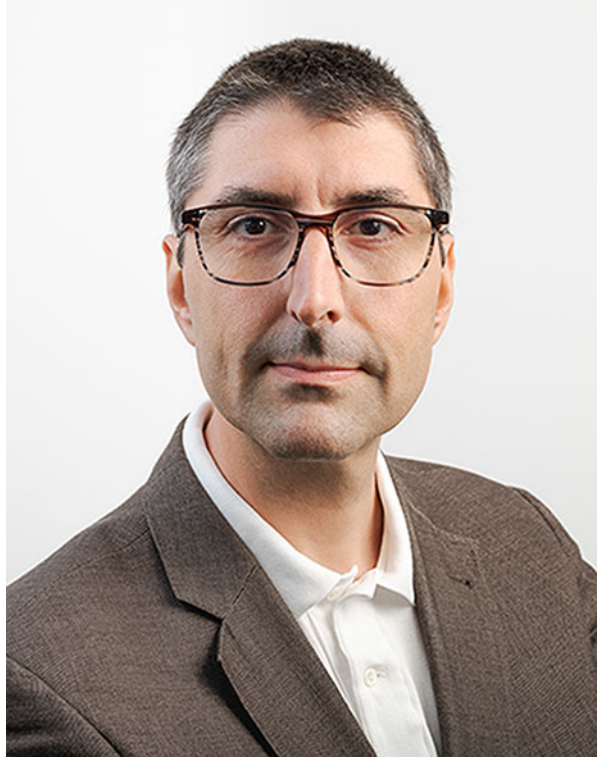

**Riccardo Scandariato** received his Ph.D. in Computer Science in 2004 from Politecnico di Torino, Italy. In his academic career, he had the opportunity to work in several countries, including the USA (University of Virginia, 2003), Italy (Politecnico di Torino, 2004–2005), Belgium (KU Leuven, 2006–2014) and Sweden (University of Gothenburg, 2014–2020). Since late 2020, he is the head of the Institute of Software Security at the Hamburg University of Technology (TUHH), in Germany. His research interests are in the field of secure software engineering, with focus on the design of secure and privacy-friendly applications.

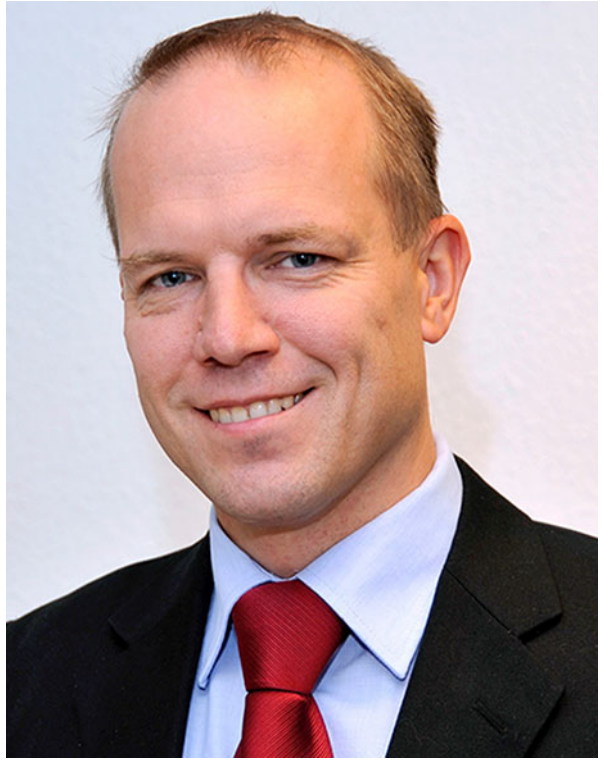

**Jan Jürjens** is a Professor, leading the Institute for Software Technology IST within the Faculty for Computer Science of the University Koblenz-Landau (Koblenz, Germany) where he is Vice Dean of Research. He is also Director Research Projects at the Fraunhofer Institute for Software and Systems Engineering ISST (Dortmund, Germany). Previous positions include a Professorship for Software Engineering at TU Dortmund, a Royal Society Industrial Fellowship at Microsoft Research Cambridge, a non-stipendiary Research Fellowship at Robinson College (Univ. Cambridge), where in 2009 he was appointed as Senior Member, and a Postdoc position at TU München. Jan holds a Doctor of Philosophy in Computing from University of Oxford and is author of "Secure Systems Development with UML" (Springer, 2005; Chinese translation 2009) and other publications mostly on software engineering and IT security. More information: http://jan.jurjens.de.